\documentclass{lmcs} 
\usepackage[utf8]{inputenc}
\pdfoutput=1

\usepackage{lastpage}
\lmcsdoi{18}{3}{16}
\lmcsheading{}{\pageref{LastPage}}{}{}%
{Jun.~15,~2021}{Aug.~02,~2022}{}


\usepackage{hyperref}

\usepackage{caption}
\usepackage{amssymb}

\newcommand\stam[1]{}
\usepackage{xspace}
\newcommand{\buchi}{B\"uchi\xspace}
\newcommand{\A}{\mathcal{A}}
\newcommand{\B}{\mathcal{B}}
\newcommand{\C}{\mathcal{C}}
\newcommand{\D}{\mathcal{D}}
\newcommand{\E}{\mathcal{E}}
\renewcommand{\S}{\mathcal{S}}

\newcommand{\zug}[1]{\langle#1\rangle}

\begin{document}

	\title[Minimization and Canonization of GFG Transition-Based Automata]{Minimization and Canonization of \texorpdfstring{\\}{}GFG Transition-Based Automata\rsuper*}
	\titlecomment{{\lsuper*} This is a full and combined version of publications~\cite{AK19} and~\cite{AK20b}.}

	\author[B.~Abu Radi]{Bader Abu Radi\lmcsorcid{0000-0001-8138-9406}} 
	\author[O.~Kupferman]{Orna Kupferman\lmcsorcid{0000-0003-4699-6117}} 
    \address{School of Computer Science and Engineering, The Hebrew University, Jerusalem, Israel}	
	\email{bader.aburadi@gmail.com, orna@cs.huji.ac.il}





	\begin{abstract}
		While many applications of automata in formal methods can use nondeterministic automata, some applications, most notably synthesis, need deterministic  or \emph{good-for-games} (GFG) automata. The latter are nondeterministic automata that can resolve their nondeterministic choices in a way that only depends on the past.
		The \emph{minimization} problem for deterministic B\"uchi and co-B\"uchi word automata is NP-complete.
		In particular, no \emph{canonical} minimal deterministic automaton exists, and a language may have
		different minimal deterministic automata.
		We describe a polynomial minimization algorithm for GFG co-B\"uchi word automata with \emph{transition-based} acceptance. Thus, a run is accepting if it traverses a set $\alpha$ of designated transitions only finitely often.
		Our algorithm is based on a sequence of transformations we apply to the automaton, on top of which a minimal quotient automaton is defined.
		We use our minimization algorithm to show canonicity for transition-based GFG co-B\"uchi word automata: all minimal automata have isomorphic safe components (namely components obtained by restricting the transitions to these not in $\alpha$) and once we saturate the automata with $\alpha$-transitions, we get full isomorphism.
	\end{abstract}

	\maketitle

	\tableofcontents

	\section{Introduction}%
	\label{intro}

	Automata theory is one of the longest established areas in Computer Science. A classical problem in automata theory is \emph{minimization}: generation of an equivalent automaton with a minimal number of states. For automata on finite words, the picture is well understood: For nondeterministic automata, minimization is PSPACE-complete~\cite{JR93}, whereas for deterministic automata, a minimization algorithm, based on the Myhill-Nerode right congruence~\cite{Myh57,Ner58}, generates in polynomial time a canonical minimal deterministic automaton~\cite{Hop71}. Essentially, the canonical automaton, a.k.a.~the \emph{quotient automaton}, is obtained by merging equivalent states.

	A prime application of automata theory is specification, verification, and synthesis of reactive systems~\cite{VW94,EKV21}. The automata-theoretic approach considers relationships between systems and their
	specifications as relationships between languages. Since we care about the on-going behavior of nonterminating  systems, the automata run on infinite words. Acceptance in such automata is determined according to the set of states that are visited infinitely often along the run. In B\"uchi automata~\cite{Buc62} (NBW and DBW, for nondeterministic and deterministic B\"uchi word automata, respectively), the acceptance condition is a subset $\alpha$ of states, and a run is accepting iff it visits $\alpha$ infinitely often. Dually, in co-B\"uchi automata (NCW and DCW), a run is accepting iff it visits $\alpha$ only finitely often.
	In spite of the extensive use of automata on infinite words in verification and synthesis algorithms and tools, some fundamental problems around their minimization are still open.
	For nondeterministic automata, minimization is PSPACE-complete, as it is for automata on finite words. Before we describe the situation for deterministic automata, let us elaborate some more on the power of  nondeterminism in the context of automata on infinite words,  as this is relevant to our contribution.

	For automata on finite words, nondeterminism does not increase the expressive power,  yet it leads to exponential succinctness~\cite{RS59}.  For automata on infinite words, nondeterminism may increase the expressive power and also leads to exponential succinctness. For example, NBWs are strictly more expressive than DBWs~\cite{Lan69}. In some applications of automata on infinite words, such as model checking, algorithms can proceed with nondeterministic automata, whereas in other applications, such as synthesis and control, they cannot. There, the advantages of nondeterminism are lost, and the algorithms involve complicated determinization constructions~\cite{Saf88} or acrobatics for circumventing determinization~\cite{KV05c}.
	Essentially, the inherent difficulty of using nondeterminism in synthesis lies in the fact that each guess of the nondeterministic automaton should accommodate all possible futures.

	A study of nondeterministic automata that can resolve their nondeterministic choices in a way that only depends on the past started in~\cite{KSV06}, where the setting is modeled by means of tree automata for derived languages. It then continued by means of  \emph{good for games}  (GFG) automata~\cite{HP06}.\footnote{GFGness is also used in~\cite{Col09} in the framework of cost functions under the name ``history-determinism''.} A nondeterministic automaton $\A$ over an alphabet $\Sigma$ is GFG if there is a strategy $g$ that maps each finite word $u \in \Sigma^*$ to the transition to be taken after $u$ is read; and following $g$ results in accepting all the words in the language of $\A$. Note that a state $q$ of $\A$ may be reachable via different words, and $g$ may suggest different transitions from $q$ after different words are read. Still, $g$ depends only on the past, namely on the word read so far. Obviously, there exist GFG automata: deterministic ones, or nondeterministic ones that are \emph{determinizable by pruning} (DBP); that is, ones that just add transitions on top of a deterministic automaton. In fact, the GFG automata constructed in~\cite{HP06} are DBP.\footnote{As explained in~\cite{HP06}, the fact that the GFG automata constructed there are DBP does not contradict their usefulness in practice, as their transition relation is simpler than the one of the embodied deterministic automaton and it can be defined symbolically.}

	In terms of expressive power, it is shown in~\cite{KSV06,NW98} that GFG automata with an acceptance condition $\gamma$ (e.g., B\"uchi) are as expressive as deterministic $\gamma$ automata. The picture in terms of succinctness is diverse. For automata on finite words, GFG automata are always DBP~\cite{KSV06,Mor03}. For automata on infinite words, in particular NBWs and NCWs, GFG automata need not be DBP~\cite{BKKS13}.
	Moreover, the best known determinization construction for GFG-NBWs is quadratic, whereas determinization of GFG-NCWs has  a tight exponential (blow up) bound~\cite{KS15}.  Thus, GFG automata on infinite words are more succinct (possibly even exponentially) than deterministic ones.\footnote{We note that some of the succinctness results are known only for GFG automata with \emph{transition-based} acceptance.}
	Further research studies characterization, typeness, complementation, and further constructions and decision procedures for GFG automata~\cite{KS15,BKS17,BK18}, as well as an extension of the GFG setting to pushdown $\omega$-automata~\cite{LZ20} and to alternating automata~\cite{BL19,BKLS20}.

	Back to the minimization problem. Recall that for finite words, an equivalent minimal deterministic automaton can be obtained by merging equivalent states. A similar algorithm is valid for determinisitic \emph{weak} automata on infinite words: DBWs in which each strongly connected component is either contained in $\alpha$ or is disjoint from $\alpha$~\cite{MSS88,Lod01}. For general DBWs (and hence, also DCWs, as the two dualize each other), merging of equivalent states fails, and minimization is NP-complete~\cite{Sch10}.

	The intractability of the minimization problem has led to a development of numerous heuristics. The heuristics either relax the minimality requirement, for example algorithms based on \emph{fair bisimulation}~\cite{GBS02}, which reduce the state space but need not return a minimal automaton, or relax the equivalence requirement, for example algorithms based on \emph{hyper-minimization}~\cite{BGS09,JM13}  or \emph{almost-equivalence}~\cite{Sch10}, which come with a guarantee about the difference between the language of the original automaton and the ones generated by the algorithm. In some cases, these algorithms do generate a minimal equivalent automaton (in particular, applying relative minimization based on almost-equivalence on a deterministic weak automaton results in an equivalent minimal weak automaton~\cite{Sch10}), but in general, they are only heuristics.
	In an orthogonal line of work, researchers have studied minimization in richer settings of automata on finite words.
	One direction is to allow some nondeterminism. As it turns out, however, even the slightest extension of the deterministic model towards a nondeterministic one, for example by allowing at most one nondeterministic choice in every accepting computation or allowing just two initial states instead of one, results in NP-complete minimization problems~\cite{Mal04}. Another direction is a study of quantitative settings. Here, the picture is diverse. For example, minimization of deterministic lattice automata~\cite{KL07} is polynomial for automata over linear lattices and is NP-complete for general lattices~\cite{HK15}, and minimization of deterministic weighted automata over the tropical semiring is polynomial~\cite{Moh97}, yet the problem is open for general semirings.

	Proving NP-hardness for DBW minimization, Schewe used a reduction from the vertex-cover problem~\cite{Sch10}.
	Essentially\footnote{The exact reduction is more complicated and involves an additional letter that is required for cases in which vertices in the graph have similar neighbours. Since, however, the vertex-cover problem is NP-hard also for instances in which different vertices have different sets of successors, it is possible to simplify this component of the reduction.}, given a graph $G=\zug{V,E}$, we seek a minimal DBW for the language $L_G$ of words of the form $v_{i_1}^+ \cdot v_{i_2}^+ \cdot v_{i_3}^+ \cdots \in V^\omega$, where for all $j \geq 1$, we have that $\zug{v_{i_j},v_{i_{j+1}}} \in E$. We can recognize $L_G$ by an automaton obtained from $G$ by adding self loops to all vertices, labelling each edge by its destination, and requiring a run to traverse infinitely many original edges of $G$. Indeed, such runs correspond to words that traverse an infinite path in $G$, possibly looping at vertices, but not getting trapped in a self loop, as required by $L_G$. When, however, the acceptance condition is defined by a set of vertices, rather than transitions, we need to duplicate some states, and a minimal duplication corresponds to a minimal vertex cover. Thus, a natural question arises: Is there a polynomial minimization algorithms for DBWs and DCWs whose acceptance condition is \emph{transition based}? In such automata, the acceptance is a subset $\alpha$ of transitions, and a run is required to traverse transitions in $\alpha$ infinitely often (in B\"uchi automata, denoted tNBW), or finitely often (in co-B\"uchi automata, denoted tNCW).
	Beyond the theoretical interest, there is recently growing use of transition-based automata in practical applications, with evidences they offer a simpler translation of LTL formulas to  automata and enable simpler constructions and decision procedures~\cite{GO01,GL02,DLFMRX16,SEJK16,LKH17}.

	In this paper we describe a polynomial-time algorithm for the minimization of GFG-tNCWs. Consider a GFG-tNCW $\A$. Our algorithm is based on a chain of transformations we apply to $\A$. Some of the transformations are introduced in~\cite{KS15}, in algorithms for deciding GFGness. These transformations result in \emph{nice} GFG-tNCWs, and satisfy some simple syntactic and semantic properties.
	We add two more transformations and prove that they guarantee minimality. Our reasoning is based on a careful analysis of the \emph{safe components} of $\A$,  namely its strongly connected components obtained by removing transitions in $\alpha$. Note that every accepting run of $\A$ eventually reaches and stays forever in a safe component. We show that a minimal GFG-tNCW equivalent to $\A$ can be obtained by defining an order on the safe components, and applying the quotient construction on a GFG-tNCW obtained by restricting attention to states that belong to components that form a frontier in this order. All involved transformations require polynomial time, thus so does the obtained minimization algorithm.

	Minimization and its complexity are tightly related to the canonicity question. Recall that $\omega$-regular languages do not have a unique minimal DBW or DCW~\cite{Kup15}.
	We continue and study canonicity for GFG and transition-based automata. We start with GFG-tNCWs and show that all minimal nice GFG-tNCWs are \emph{safe isomorphic}, namely their safe components are isomorphic. More formally, if $\A_1$ and $\A_2$ are nice minimal GFG-tNCWs for the same language, then there exists a bijection between the state spaces of $\A_1$ and $\A_2$ that induces a bijection between their $\bar{\alpha}$-transitions (these not in $\alpha$).  We then show that by saturating the GFG-tNCW with $\alpha$-transitions we get isomorphism among all minimal automata.
	We suggest two possible saturations.
	One adds as many $\alpha$-transitions as possible, and the second does so in a way that preserves $\alpha$-homogeneity, thus for every state $q$ and letter $\sigma$, all the transitions labeled $\sigma$ from $q$ are $\alpha$-transitions or are all
	$\bar{\alpha}$-transitions. Since our minimization algorithm generates minimal $\alpha$-homogenous GFG-tNCWs, both forms of canonical minimal GFG-tNCWs can be obtained in polynomial time.

	We then show that GFGness is not a sufficient condition for canonicity, raising the question of canonicity in tDCWs. Note that unlike the GFG-tNCW setting, dualization of the acceptance condition in deterministic automata complements the language of an automaton, and thus our results apply also to canonicity of tDBWs. We start with some bad news, showing that as has been the case with DCWs and DBWs, minimal tDCWs and tDBWs need not be isomorphic. Moreover, we cannot change membership of transitions in $\alpha$ or add transitions in order to make them isomorphic. On the positive side, safe isomorphism is helpful also in the tDCW setting: Consider an $\omega$-regular language $L$. Recall that the minimal  GFG-tNCW for $L$ may be smaller than a minimal tDCW for $L$~\cite{KS15}. We say that $L$ is \emph{GFG-helpful} if this is the case. We prove that all minimal tDCWs for an $\omega$-regular language that is not GFG-helpful are safe isomorphic.
	For $\omega$-regular languages that are GFG-helpful, safe isomorphism is left open. For such languages, however, we care more about minimal GFG-tNCWs, which may be exponentially smaller~\cite{KS15}, and which do have a canonical form.  Also, natural $\omega$-regular languages are not GFG-helpful, and in fact the existence of $\omega$-regular languages that are GFG-helpful had been open for quite a while~\cite{BKKS13}. Accordingly, we view our results as good news about canonicity in deterministic automata with transition-based acceptance.

	\section{Preliminaries}%
	\label{prelim}

	For a finite nonempty alphabet $\Sigma$, an infinite \emph{word} $w = \sigma_1 \cdot \sigma_2 \cdots \in \Sigma^\omega$ is an infinite sequence of letters from $\Sigma$.
	A \emph{language} $L\subseteq \Sigma^\omega$ is a set of words. We denote the empty word by $\epsilon$, and the set of finite words over $\Sigma$ by $\Sigma^*$. For $i\geq 0$, we use $w[1, i]$ to denote the (possibly empty) prefix $\sigma_1\cdot \sigma_2 \cdots  \sigma_i$ of $w$ and use $w[i+1, \infty]$ to denote its suffix $\sigma_{i+1} \cdot  \sigma_{i+2} \cdots$.

	A \emph{nondeterministic automaton} over infinite words is $\A = \langle \Sigma, Q, q_0, \delta, \alpha  \rangle$, where $\Sigma$ is an alphabet, $Q$ is a finite set of \emph{states}, $q_0\in Q$ is an \emph{initial state}, $\delta: Q\times \Sigma \to 2^Q\setminus \emptyset$ is a \emph{transition function}, and $\alpha$ is an \emph{acceptance condition}, to be defined below. For states $q$ and $s$ and a letter $\sigma \in \Sigma$, we say that $s$ is a $\sigma$-successor of $q$ if $s \in \delta(q,\sigma)$.
	Note that $\A$ is \emph{total}, in the sense that it has at least one successor for each state and letter, and that $\A$ may be \emph{nondeterministic}, as the transition function may specify several successors for each state and letter.
	If $|\delta(q, \sigma)| = 1$ for every state $q\in Q$ and letter $\sigma \in \Sigma$, then $\A$ is \emph{deterministic}.
	We define the \emph{size} of $\A$, denoted $|\A|$, as its number of states\footnote{Note that our definition of size follows the traditional complexity measure of deterministic automata on finite words. As argued in~\cite{Bok17}, it is interesting to study additional complexity measures, in particular the size of the transition function of the automaton.}, thus, $|\A| = |Q|$.

	When $\A$ runs on an input word, it starts in the initial state and proceeds according to the transition function. Formally, a \emph{run}  of $\A$ on $w = \sigma_1 \cdot \sigma_2 \cdots \in \Sigma^\omega$ is an infinite sequence of states $r = r_0,r_1,r_2,\ldots \in Q^\omega$, such that $r_0 = q_0$, and for all $i \geq 0$, we have that $r_{i+1} \in \delta(r_i, \sigma_{i+1})$.
	We sometimes consider finite runs on finite words. In particular,
	we sometimes extend $\delta$ to sets of states and finite words. Then, $\delta: 2^Q\times \Sigma^* \to 2^Q$ is such that for every $S \in 2^Q$, finite word $u\in \Sigma^*$, and letter $\sigma\in \Sigma$, we have that $\delta(S, \epsilon) = S$, $\delta(S, \sigma) = \bigcup_{s\in S}\delta(s, \sigma)$, and $\delta(S, u \cdot \sigma) = \delta(\delta(S, u), \sigma)$. Thus, $\delta(S, u)$ is the set of states that $\A$ may reach when it reads $u$ from some state in $S$.

	The transition function $\delta$ induces a transition relation $\Delta \subseteq Q\times \Sigma \times Q$, where for every two states $q,s\in Q$ and letter $\sigma\in \Sigma$, we have that $\langle q, \sigma, s \rangle \in \Delta$ iff $s\in \delta(q, \sigma)$.
	We sometimes view the run $r = r_0,r_1,r_2,\ldots$ on $w = \sigma_1 \cdot \sigma_2 \cdots$ as an infinite sequence of successive transitions $\zug{r_0,\sigma_1,r_1}, \zug{r_1,\sigma_2,r_2},\ldots \in \Delta^\omega$.
	The acceptance condition $\alpha$ determines which runs are ``good''. We consider here \emph{transition-based} automata, in which $\alpha$ is a set of transitions --- these that should be traversed infinitely often during the run; specifically, $\alpha\subseteq \Delta$. We use the terms \emph{$\alpha$-transitions} and  \emph{$\bar{\alpha}$-transitions} to refer to  transitions in $\alpha$ and in $\Delta \setminus \alpha$, respectively. We also refer to restrictions $\delta^\alpha$ and  $\delta^{\bar{\alpha}}$ of $\delta$, where for all $q,s \in Q$ and $\sigma \in \Sigma$, we have that $s \in \delta^\alpha(q, \sigma)$ iff $\zug{q,\sigma,s} \in \alpha$, and $s \in \delta^{\bar{\alpha}}(q, \sigma)$ iff $\zug{q,\sigma,s} \in \Delta \setminus \alpha$.

	We say that $\A$ is \emph{$\alpha$-homogenous} if for every state $q\in Q$ and letter $\sigma \in \Sigma$, either $\delta^\alpha_\A(q, \sigma) =\emptyset$ or $\delta^{\bar{\alpha}}_\A(q, \sigma) = \emptyset$. Thus, either all the $\sigma$-labeled transitions from $q$ are $\alpha$-transitions, or they are all $\bar{\alpha}$-transitions.
	For a run $r \in \Delta^\omega$, let ${\textit{inf}}(r)\subseteq \Delta$ be the set of transitions that $r$ traverses infinitely often. Thus,
	${\it inf}(r) = \{  \langle q, \sigma, s\rangle \in \Delta: q = r_i, \sigma = \sigma_{i+1} \text{ and } s = r_{i+1} \text{ for infinitely many $i$'s}   \}$.
	In \emph{co-B\"uchi} automata, $r$ is \emph{accepting} iff ${\it inf}(r)\cap \alpha = \emptyset$, thus if $r$ traverses transitions in $\alpha$ only finitely often. A run that is not accepting is \emph{rejecting}.  A word $w$ is accepted by $\A$ if there is an accepting run of $\A$ on $w$. The language of $\A$, denoted $L(\A)$, is the set of words that $\A$ accepts. Two automata are \emph{equivalent} if their languages are equivalent. We use tNCW and tDCW to abbreviate nondeterministic and deterministic transition-based co-B\"uchi automata over infinite words, respectively.
	We say that a nondeterministic automaton $\A$ is \emph{determinizable by prunning} (DBP) if we can remove some of the transitions of $\A$ and get a deterministic automaton that recognizes $L(\A)$.

	We continue to definitions and notations that are relevant to our study. See Section~\ref{app glos} for a glossary.
	For a state $q\in Q$ of an automaton $\A = \langle \Sigma, Q, q_0, \delta, \alpha \rangle$, we define $\A^q$ to be the automaton obtained from $\A$ by setting the initial state to be $q$. Thus, $\A^q = \langle \Sigma, Q, q, \delta, \alpha \rangle$.
	We say that two states $q,s\in Q$ are \emph{equivalent}, denoted $q \sim_{\A} s$, if $L(\A^q) = L(\A^s)$.
	The automaton $\A$ is \emph{semantically deterministic} if different nondeterministic choices lead to equivalent states. Thus, for every state $q\in Q$ and letter $\sigma \in \Sigma$, all the $\sigma$-successors of $q$ are equivalent: for every two states $s, s'\in Q$ such that  $\langle q, \sigma, s\rangle$ and $\langle q, \sigma, s'\rangle$ are in $\Delta$, we have that $s \sim_{\A} s'$.
	The following proposition follows immediately from the definitions.

	\begin{prop}\label{pruned-corollary}
		Consider a semantically deterministic automaton $\A$, states $q,s \in Q$, letter $\sigma\in \Sigma$, and transitions $\langle q, \sigma, q'\rangle,\langle s, \sigma, s'\rangle \in \Delta$. If $q \sim_{\A} s$, then $q' \sim_{\A} s'$.
	\end{prop}

	An automaton $\A$ is \emph{good for games} (\emph{GFG}, for short) if its nondeterminism can be resolved based on the past, thus on the prefix of the input word read so far. Formally, $\A$ is \emph{GFG} if there exists a \emph{strategy} $f:\Sigma^* \to Q$ such that the following hold:
	\begin{enumerate}
		\item
		The strategy $f$ is consistent with the transition function. That is, $f(\epsilon)=q_0$, and for every finite word $u \in \Sigma^*$ and letter $\sigma \in \Sigma$, we have that $\zug{f(u),\sigma,f(u \cdot \sigma)} \in \Delta$.
		\item
		Following $f$ causes $\A$ to accept all the words in its language. That is, for every infinite word $w = \sigma_1 \cdot \sigma_2 \cdots \in \Sigma^\omega$, if $w \in L(\A)$, then the run $f(w[1, 0]), f(w[1, 1]), f(w[1, 2]), \ldots$, which we denote by $f(w)$, is 
		an accepting run of $\A$ on $w$.
	\end{enumerate}
	Note that by definition, the strategy $f$ maps a finite word $u$ to the state that the run induced by $f$ visits after reading $u$. We sometimes abuse notation and use $f(w)$, for an infinite word $w$, to denote the run on $w$  that is  induced by $f$.
	We say that the strategy $f$ \emph{witnesses} $\A$'s GFGness.
	For an automaton $\A$, we say that a state $q$ of $\A$ is \emph{GFG} if $\A^q$ is GFG\@.
	Note that every deterministic automaton is GFG\@.

	Consider a directed graph $G = \langle V, E\rangle$. A \emph{strongly connected set} in $G$ (SCS, for short) is a set $C\subseteq V$ such that for every two vertices $v, v'\in C$, there is a path from $v$ to $v'$. A SCS is \emph{maximal} if it is maximal w.r.t containment, that is, for every non-empty set $C'\subseteq V\setminus C$, it holds that $C\cup C'$ is not a SCS\@. The \emph{maximal strongly connected sets} are also termed \emph{strongly connected components} (SCCs, for short). The \emph{SCC graph of $G$} is the graph defined over the SCCs of $G$, where there is an edge from a SCC $C$ to another SCC $C'$ iff there are two vertices $v\in C$ and $v'\in C'$ with $\langle v, v'\rangle\in E$. A SCC is \emph{ergodic} iff it has no outgoing edges in the SCC graph. The SCC graph of $G$ can be computed in linear time by standard SCC algorithms~\cite{Tar72}.
	An automaton $\A = \langle \Sigma, Q, q_0, \delta, \alpha\rangle$ induces a directed graph $G_{\A} = \langle Q, E\rangle$, where $\langle q, q'\rangle\in E$ iff there is a letter $\sigma \in \Sigma$ such that $\langle q, \sigma, q'\rangle \in \Delta$. The SCSs and SCCs of $\A$ are those of $G_{\A}$.
	We say that a tNCW $\A$ is \emph{safe deterministic} if by removing its $\alpha$-transitions, we remove all nondeterministic choices. Thus, for every state $q\in Q$ and letter $\sigma\in \Sigma$, it holds that $|\delta^{\bar{\alpha}}(q, \sigma)|\leq 1$.

	We refer to the SCCs we get by removing $\A$'s $\alpha$-transitions as the \emph{safe components} of $\A$; that is, the \emph{safe components} of $\A$ are the SCCs of the graph $G_{\A^{\bar{\alpha}}} = \langle Q, E^{\bar{\alpha}} \rangle$, where $\zug{q, q'}\in E^{\bar{\alpha}}$ iff there is a letter $\sigma\in \Sigma$ such that $\langle q, \sigma, q'\rangle \in \Delta \setminus \alpha$. We denote the set of safe components of $\A$ by $\S(\A)$. For a safe component $S\in \S(\A)$, the \emph{size} of $S$, denoted $|S|$, is the number of states in $S$.
	Note that an accepting run of $\A$ eventually gets trapped in one of $\A$'s safe components. A tNCW $\A$ is \emph{normal} if
	there are no $\bar{\alpha}$-transitions connecting different safe components. That is,
	for all states $q$ and $s$ of $\A$, if there is a path of $\bar{\alpha}$-transitions from $q$ to $s$, then there is also a path of $\bar{\alpha}$-transitions from $s$ to $q$.

	We now combine several properties defined above and say that a GFG-tNCW $\A$ is \emph{nice} if all the states in $\A$ are reachable and GFG, and $\A$ is normal, safe deterministic, and semantically deterministic.
	In the theorem below we combine arguments from~\cite{KS15} showing that each of these properties can be obtained in at most polynomial time, and without the properties being conflicting. For some properties, we give an alternative and simpler proof.
	\begin{thmC}[\cite{KS15}]\label{nice}
		Every GFG-tNCW $\A$ can be turned, in polynomial time, into an equivalent nice GFG-tNCW $\B$ such that $|\B|\leq |\A|$.
	\end{thmC}

	\begin{proof}
		It is shown in~\cite{KS15} that one can decide the GFGness of a tNCW $\A$ in polynomial time. The proof goes through an intermediate step where the authors construct a two-player game such that if the first player does not win the game, then $\A$ is not GFG, and otherwise a winning strategy for the first player induces
		a safe deterministic GFG-tNCW $\B$ equivalent to $\A$  of size at most $|\A|$. As we start with a GFG-tNCW $\A$, such a winning strategy is guaranteed to exist, and, by~\cite{KS15}, can be found in polynomial time. Thus, we obtain an equivalent safe deterministic GFG-tNCW $\B$ of size at most $|\A|$  in polynomial time. In fact, it can be shown that $\B$ is also semantically deterministic. Yet, for completeness we give below a general procedure for semantic determinization.

		For a tNCW $\A$, we say that a transition $\langle q, \sigma, s \rangle \in \Delta$ is \emph{covering} if for every transition $\langle q, \sigma, s'\rangle$, it holds that $L({\A}^{s'}) \subseteq L(\A^s)$.  If $\A$ is GFG and $f$ is a strategy witnessing its GFGness, we say that a state $q$ of $\A$ is \emph{used by} $f$ if there is a finite word $u$ with $f(u) = q$, and we say that a transition $\langle q, \sigma, q'\rangle$ of $\A$ is \emph{used by} $f$ if there is a finite word $u$ with $f(u) = q$ and $f(u\cdot\sigma) = q'$.
		Since states that are not GFG can be detected in polynomial time, and as all states that are used by a strategy that witnesses $\B$'s GFGness are GFG, the removal of non-GFG states does not affect $\B$'s language. Note that removing the non-GFG states may result in a non-total automaton, in which case we add a rejecting sink. Note that we add a rejecting sink only after removing some state, thus this step does not add states.
		Now, using the fact that language containment of GFG-tNCWs can be checked in polynomial time~\cite{HKR02,KS15}, and transitions that are used by strategies are covering~\cite{KS15}, one can semantically determinize $\B$ by removing non-covering transitions.%

		States that are not reachable are easy to detect, and their removal does not affect $\B$'s language nor its GFGness. Normalization is also easy to obtain and involves adding some existing transitions to $\alpha$~\cite{KS15}.
		Indeed,  if there are $\bar{\alpha}$-transitions connecting different safe components of $\B$, then they can be added to $\alpha$ without affecting the acceptance of runs in $\B$, as every accepting run traverses such transitions only finitely often. Thus, the language and GFGness of all states are not affected.
		Finally, it is not hard to verify that the properties, in the order we obtain them in the proof, are not conflicting, and thus the described sequence of transformations results in a nice GFG-tNCW whose size is bounded by $|\A|$.
	\end{proof}

	\section{Minimizing GFG-tNCWs}\label{sc minimization}
	A GFG-tNCW $\A$ is \emph{minimal} if for every equivalent GFG-tNCW $\B$, it holds that $|\A|\leq |\B|$.
	In this section, we suggest a polynomial-time minimization algorithm for GFG-tNCWs.

	\subsection{A Sufficient Condition for GFG-tNCW Minimality}\label{s condition}

	In this section, we define two additional properties for nice GFG-tNCWs, namely \emph{safe-centralized} and \emph{safe-minimal}, and we prove that nice GFG-tNCWs that satisfy these properties are minimal. In Sections~\ref{sec sc} --~\ref{sec sd and sm}, we are going to show that the two properties can be attained in polynomial time. Before we start, let us note that a GFG-tNCW may be nice and still not be minimal. A simple example is the tDCW $\A_{\sf fm}$ for the language $(a+b)^* \cdot a^\omega$ that appears in Figure~\ref{fig afm} below. The dashed transitions are $\alpha$-transitions. It is easy to see that $\A_{\sf fm}$ is a nice GFG-tNCW, but it is not minimal.
	\begin{figure}[htb]
		\begin{center}
			\includegraphics[width=.35\textwidth]{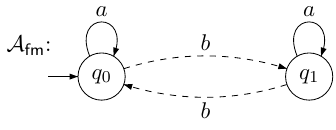}
			\caption{The tDCW $\A_{\sf fm}$.}%
			\label{fig afm}
		\end{center}
	\end{figure}

	Consider a tNCW $\A = \langle \Sigma, Q, q_0, \delta, \alpha \rangle$.
	A run $r$ of $\A$ is \emph{safe} if it does not traverse $\alpha$-transitions.
	The \emph{safe language} of $\A$, denoted $L_{{\it safe}}(\A)$, is the set of infinite words $w$, such that there is a safe run of $\A$ on $w$.
	Accordingly, each state $q \in Q$ induces two languages:  the language of $q$, namely $L(\A^q)$, which includes words accepted from $q$, and the safe language of $q$, namely $L_{{\it safe}}(\A^q)$, which is the set of infinite words that can be read from $q$ without traversing an $\alpha$-transition. Clearly, $L_{{\it safe}}(\A^q) \subseteq L(\A^q)$.
	Recall that two states $q,s\in Q$ are equivalent ($q \sim_{\A} s$) if $L(\A^q) = L(\A^s)$.
	Then, $q$ and $s$ are  \emph{strongly-equivalent}, denoted $q \approx _{\A} s$, if $q \sim_{\A} s$ and $L_{{\it safe}}(\A^q) = L_{{\it safe}}(\A^s)$. Thus, $q$ and $s$ agree on both their languages and safe languages. Finally, $q$ is \emph{subsafe-equivalent to} $s$, denoted $q\precsim_{\A} s$, if $q \sim_{\A} s$ and $L_{{\it safe}}(\A^q) \subseteq L_{{\it safe}}(\A^s)$. Note that the three relations are transitive. When $\A$ is clear from the context, we omit it from the notations, thus write 
	$q\precsim s$, $q\approx s$, etc.
	The tNCW $\A$ is \emph{safe-minimal} if it has no strongly-equivalent states. Thus, every two states differ in their languages or safe-languages.
	Then, $\A$ is \emph{safe-centralized} if for every two states $q, s\in Q$, if $q \precsim s$, then $q$ and $s$ are in the same safe component of $\A$. Intuitively, this suggests that $\A$ is compact, in the sense that states that agree on their languages and have comparable safe languages are not spread in different safe components.

	\begin{exa}
		{\rm The nice GFG-tNCW $\A_{\sf fm}$ from Figure~\ref{fig afm} is neither safe-minimal (its two states are strongly-equivalent) nor safe-centralized (its two states are in different safe components). As another example, consider the tDCW $\A$ appearing in Figure~\ref{safe example}. Recall that the dashed transitions are $\alpha$-transitions.
			All the states of $\A$ are equivalent, yet they all differ in their safe language. Accordingly, $\A$ is safe-minimal.
			Since $\{a^\omega\}= L_{{\it safe}}(\A^{q_2}) \subseteq L_{{\it safe}}(\A^{q_0})$, we have that $q_2 \precsim q_0$. Hence, as $q_0$ and $q_2$ are in different safe components, the tDCW $\A$ is not safe-centralized. \hfill \qed}

		\begin{figure}[htb]
			\begin{center}
				\includegraphics[width=.35\textwidth]{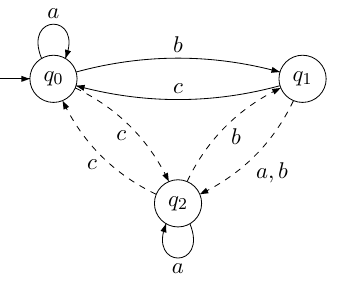}
				\caption{The tDCW $\A$.}%
				\label{safe example}
			\end{center}
		\end{figure}
	\end{exa}

We continue and specify useful properties of nice GFG-tNCWs. Intuitively, these properties show that the coincidence of language equivalence with \emph{bisimulation}~\cite{Mil71} in the setting of deterministic automata is carried over to GFG-tNCWs when restricted to their safe components.

	\begin{prop}\label{equiv-to-equiv}
		Consider a nice GFG-tNCW $\A$ and states $q$ and $s$ of $\A$ such that
		$q \approx s$ ($q \precsim s$). For every letter $\sigma \in \Sigma$ and $\bar{\alpha}$-transition $\langle q, \sigma, q'\rangle$, there is an $\bar{\alpha}$-transition $\langle s, \sigma, s'\rangle$ such that $q' \approx s'$ ($q' \precsim s'$, respectively).
	\end{prop}

	\begin{proof}
		We prove the proposition for the case $q \approx s$. The case $q \precsim s$ is similar.
		Since $\A$ is normal, the existence of the $\bar{\alpha}$-transition $\langle q, \sigma, q'\rangle$ implies that there is a safe run from $q'$ back to $q$. Hence, there is a word $z \in L_{{\it safe}}(\A^{q'})$. Clearly, $\sigma \cdot z$ is in $L_{{\it safe}}(\A^q)$. Now, since $q \approx s$, we have that $L_{{\it safe}}(\A^q) = L_{{\it safe}}(\A^s)$. In particular, $\sigma \cdot z\in L_{{\it safe}}(\A^s)$, and thus there is an $\bar{\alpha}$-transition $\langle s, \sigma, s'\rangle$. We prove that $q' \approx s'$. Since $L(\A^q) = L(\A^s)$ and $\A$ is semantically deterministic, then, by Proposition~\ref{pruned-corollary}, we have that $L(\A^{q'}) = L(\A^{s'})$. It is left to prove that $L_{{\it safe}}(\A^{q'}) = L_{{\it safe}}(\A^{s'})$. We prove that $L_{{\it safe}}(\A^{q'}) \subseteq L_{{\it safe}}(\A^{s'})$. The second direction is similar. Since $\A$ is safe deterministic, the transition $\langle s, \sigma, s'\rangle$ is the only $\sigma$-labeled $\bar{\alpha}$-transition from $s$. Hence, if by contradiction there is a word $z\in L_{{\it safe}}(\A^{q'}) \setminus L_{{\it safe}}(\A^{s'})$, we get that $\sigma \cdot z \in L_{{\it safe}}(\A^q)\setminus L_{{\it safe}}(\A^s)$, contradicting the fact that $L_{{\it safe}}(\A^q) = L_{{\it safe}}(\A^s)$.
	\end{proof}

	We continue with propositions that relate two automata, $\A=\zug{\Sigma, Q_{\A}, q^0_{\A}, \delta_{\A}, \alpha_{\A}}$ and $\B=\zug{\Sigma, Q_{\B}, q^0_{\B}, \delta_{\B}, \alpha_{\B}}$. We assume that $Q_{\A}$ and $Q_{\B}$ are disjoint, and extend the $\sim$, $\approx$, and $\precsim$ relations to states in $Q_{\A} \cup Q_{\B}$ in the expected way. For example, for $q\in Q_{\A}$ and $s\in Q_{\B}$, we use $q \sim s$ to indicate that $L(\A^q)=L(\B^s)$.
	Propositions~\ref{there-has-to-be-a-subsafe} and~\ref{there is the same safe} suggest that states of equivalent GFG-tNCWs are related through their languages and safe-languages. In Lemma~\ref{injection}, we use this fact in order to show that among equivalent GFG-tNCWs, safe-centralized automata have minimal number of safe-components.

	\begin{prop}\label{there-has-to-be-a-subsafe}
		Let $\A$ and $\B$ be equivalent nice GFG-tNCWs. For every state $q\in Q_{\A}$, there is a state $s\in Q_{\B}$ such that $q \precsim s$.
	\end{prop}

	\begin{proof}
		Let $g$ be a strategy witnessing $\B$'s GFGness. Consider a state $q\in Q_{\A}$. Let $u \in \Sigma^*$ be such that $q \in \delta_{\A}(q^0_\A, u)$. Since $\A$ and $\B$ are equivalent and semantically deterministic, an iterative application of Proposition~\ref{pruned-corollary} implies that for every state $s \in \delta_{\B}(q^0_\B, u)$, we have $q \sim s$. In particular, $q \sim g(u)$. If $L_{{\it safe}}(\A^q) = \emptyset$, then we are done, as $L_{{\it safe}}(\A^q)\subseteq L_{{\it safe}}(\B^{g(u)})$. If $L_{{\it safe}}(\A^q) \neq \emptyset$, then the proof proceeds as follows. Assume by way of contradiction that for every state $s\in Q_{\B}$ that is equivalent to $q$, it holds that $L_{{\it safe}}(\A^q)\not \subseteq L_{{\it safe}}(\B^s)$. We define an infinite word $z \in \Sigma^\omega$ such that $\A$ accepts $u \cdot z$, yet $g(u \cdot z)$ is a rejecting run of $\B$. Since $\A$ and $\B$ are equivalent, this contradicts the fact that $g$ witnesses $\B$'s GFGness.

		Before defining the word $z$, note that the following holds for every state $s\in Q_\B$ that is equivalent to $q$. Recall that we assume that $L_{{\it safe}}(\A^q)\not \subseteq L_{{\it safe}}(\B^{s})$. Hence, there is a finite nonempty word $z_s \in \Sigma^*$ such that there is a safe run of $\A^q$ on $z_s$, but every run of $\B^{s}$ on $z_s$ is not safe.
		To see why, assume contrarily that for every finite nonempty word $z_s \in \Sigma^*$, if $q$ has a safe run  on $z_s$, then there is also a safe run of $\B^{s}$ on $z_s$. Then, consider an infinite word $x = \sigma_1\cdot \sigma_2 \cdot \sigma_3 \cdots \in L_{{\it safe}}(\A^q)\setminus L_{{\it safe}}(\B^s)$. For all $i\geq 1$, the state $q$ has a safe run on the prefix $x[1, i]$. Therefore, by the assumption, so does $s$. Now, as $\B$ is nice, in particular, safe deterministic, it follows that the safe runs of $s$ on different prefixes of $x$ extend each other. Hence, one can iteratively define a safe run $r = s_0, s_1, s_2, \ldots$ of $s$ on $x$ by taking $s_i$ to be the state that is reachable by reading $x[1, i]$ from $s$ via a safe run, which contradicts the fact that $x\notin L_{{\it safe}}(\B^s)$.

		Now back to our Proposition. We can define the word $z \in \Sigma^\omega$ as follows.  Let $s_{0} = g(u)$. Since $L_{{\it safe}}(\A^q)\not \subseteq L_{{\it safe}}(\B^{s_{0}})$, then as argued above, there is a finite nonempty word $z_1$ such that there is a safe run of $\A^q$ on $z_1$, but every run of $\B^{s_{0}}$ on $z_1$ is not safe.
		In particular, the run of $\B^{s_{0}}$ that is induced by $g$, namely $g(u),g(u \cdot z_1[1, 1]),g(u \cdot z_1[1, 2]),\ldots, g(u \cdot z_1)$, traverses an $\alpha$-transition. Since $\A$ is normal, we can define  $z_1$ so the safe run of $\A^q$ on $z_1$ ends in $q$. Let $s_{1}=g(u \cdot z_1)$. We have so far two finite runs: $q \xrightarrow{z_1} q$ in $\A$  and $s_{0} \xrightarrow{z_1}s_{1}$ in $\B$. The first run is safe, and the second is the non-safe run induced by $g$ on the input word $u\cdot z_1$. Now, since $q \sim s_{0}$,  then again by Proposition~\ref{pruned-corollary} we have that $q \sim s_{1}$, and by applying the same considerations, we can define a finite nonempty word $z_2$ and $s_{2}=g(u \cdot z_1 \cdot z_2)$ such that $q \xrightarrow{z_2} q$ and $s_{1} \xrightarrow{z_2}s_{2}$. Here too, the first run is safe, and the second is the non-safe run induced by $g$ on the input word $u\cdot z_1\cdot z_2$.
		 Proceeding similarly, we can iteratively define an infinite word $z = z_1\cdot z_2 \cdot z_3 \cdots $, where for all $i\geq 1$, we have that $q \xrightarrow{z_i} q$  and $s_{i-1} \xrightarrow{ z_i}s_{i} = g(u\cdot z_1\cdot z_2 \cdots z_i)$, where the first run is safe, and the second is the non-safe run induced by $g$ on the input word $u\cdot z_1 \cdot z_2 \cdots z_i$. Also, $q\sim s_i$.
		On the one hand, since $q \in \delta_{\A}(q^0_\A, u)$ and there is a safe run of $\A^q$ on $z$, we have that $u \cdot z \in L(\A)$. On the other hand, the run of $\B$ on the infinite word $u\cdot z$ that is induced by $g$, namely $g(u \cdot z)$, traverses $\alpha$-transitions infinitely often, and is thus rejecting.
	\end{proof}

	\begin{prop}\label{there is the same safe}
		Let $\A$ and $\B$ be equivalent nice GFG-tNCWs. For every state $p\in Q_{\A}$, there are states $q\in Q_{\A}$ and $s\in Q_{\B}$ such that $p \precsim q$ and $q \approx s$.
	\end{prop}

	\begin{proof}
		The proposition follows from the combination of Proposition~\ref{there-has-to-be-a-subsafe} with the transitivity of $\precsim$ and the fact $Q_\A$ and $Q_\B$ are finite. Formally, consider the directed bipartite graph $G = \langle Q_{\A} \cup Q_{\B}, E \rangle$, where $E\subseteq (Q_{\A}\times Q_{\B}) \cup (Q_{\B}\times Q_{\A})$ is such that $\langle p_1, p_2\rangle \in E$ iff  $p_1 \precsim p_2$. Proposition~\ref{there-has-to-be-a-subsafe} implies that $E$ is total. That is, from every state in $Q_{\A}$ there is an edge to some state in $Q_{\B}$, and from every state in $Q_{\B}$ there is an edge to some state in $Q_{\A}$.
		Since $Q_\A$ and $Q_\B$ are finite, this implies that for every $p \in Q_{\A}$, there is a path in $G$ that starts in $p$ and reaches a state $q\in Q_{\A}$ (possibly $q=p$) that belongs to a nonempty cycle. We take $s$ to be some state in $Q_\B$ in this cycle. By the transitivity of $\precsim$, we have that $p \precsim q$, $q \precsim s$, and $s \precsim q$. The last two imply that  $q \approx s$, and we are done.
	\end{proof}

	\begin{lem}%
		\label{injection}
		Consider a nice GFG-tNCW $\A$. If $\A$ is safe-centralized and safe-minimal, then for every nice GFG-tNCW $\B$ equivalent to $\A$, there is an injection $\eta: \S(\A) \to \S(\B)$ such that for every safe component $T\in \S(\A)$, it holds that $|T|\leq |\eta(T)|$.
	\end{lem}

	\begin{proof}
		We define $\eta$ as follows. Consider a safe component $T\in \S(\A)$.  Let $p_{T}$ be some state in $T$. By Proposition~\ref{there is the same safe}, there are states $q_{T} \in Q_{\A}$ and $s_{T} \in Q_{\B}$ such that $p_{T} \precsim q_{T}$ and $q_{T} \approx s_{T}$. Since $\A$ is safe-centralized, the states $p_{T}$ and $q_{T}$ are in the same safe component, thus $q_{T} \in T$. We define $\eta(T)$ to be the safe component of $s_{T}$ in $\B$. We show that $\eta$ is an injection; that is, for every two distinct safe components $T_1$ and $T_2$ in $\S(\A)$, it holds that $\eta(T_1)\neq \eta(T_2)$. Assume by way of contradiction that $T_1$ and $T_2$ are such that $s_{T_1}$ and $s_{T_2}$, chosen as described above, are in the same safe component of $\B$. Then, there is a safe run from $s_{T_1}$ to $s_{T_2}$. Since $s_{T_1} \approx q_{T_1}$, an iterative application of Proposition~\ref{equiv-to-equiv} implies that there is a safe run from $q_{T_1}$ to some state $q$ such that $q \approx s_{T_2}$. Since the run from $q_{T_1}$ to $q$ is safe, the states $q_{T_1}$ and $q$ are in the same safe component, and so $q \in T_1$. Since $q_{T_2} \approx s_{T_2}$, then $q \approx q_{T_2}$.
		Since $\A$ is safe-centralized, the latter implies that $q$ and $q_{T_2}$ are in the same safe component, and so $q \in T_2$, and we have reached a contradiction.

		It is left to prove that for every safe component $T\in \S(\A)$, it holds that $|T|\leq |\eta(T)|$. Let $T\in \S(\A)$ be a safe component of $\A$. By the definition of $\eta$, there are $q_{T}\in T$ and $s_{T}\in \eta(T)$ such that $q_{T} \approx s_T$. 
		 Since $T$ is a safe component,  there is a safe run $q_0,q_1,\ldots, q_m$ of $\A$ that starts in $q_T$ and traverses all the states in $T$. Since $\A$ is safe-minimal, no two states in $T$ are strongly equivalent. Therefore, there is a subset $I\subseteq \{0, 1, \ldots, m\}$ of indices, with $|I| = |T|$, such that for every two  different indices $i_1, i_2 \in I$, it holds that $q_{i_1} \not \approx q_{i_2}$.
		By applying Proposition~\ref{equiv-to-equiv} iteratively, there is a safe run $s_0,s_1,\ldots, s_m$ of $\B$ that starts in $s_T$ and such that for every $0 \leq i \leq m$, it holds that $q_i \approx s_i$. Since the run is safe and $\B$ is normal, it stays in $\eta(T)$.  Then, however, for every two different indices $i_1, i_2 \in I$, we have that $s_{i_1} \not \approx s_{i_2}$, and so $s_{i_1} \neq s_{i_2}$. Hence, $|\eta(T)|\geq |I| = |T|$.
	\end{proof}

	We can now prove that the additional two properties imply the minimality of nice GFG-tNCWs.

	\begin{thm}\label{C is minimal}
		Consider a nice GFG-tNCW $\A$. If $\A$ is safe-centralized and safe-minimal, then $\A$ is a minimal  GFG-tNCW for $L(\A)$.
	\end{thm}

	\begin{proof}
		Let $\B$ be a GFG-tNCW equivalent to $\A$. By Theorem~\ref{nice}, we can assume that $\B$ is nice. Indeed, otherwise we can make it nice without increasing its state space. Then, by Lemma~\ref{injection}, there is an injection $\eta: \S(\A) \to \S(\B)$ such that for every safe component $T\in \S(\A)$, it holds that $|T|\leq |\eta(T)|$. Hence,
		\[|\A| = \sum\limits_{T\in \S(\A)} |T| \leq \sum\limits_{T\in \S(\A)} |\eta(T)| \leq \sum\limits_{{T}' \in \S(\B)} |{T}'| =  |\B|.\] Indeed, the first inequality follows from the fact $|T| \leq |\eta(T)|$, and the second inequality follows from the fact that $\eta$ is injective.
	\end{proof}

	\begin{rem}
		{\rm Recall that we assume that the transition function of GFG-tNCWs is total. Clearly, a non-total GFG-tNCW can be made total by adding a rejecting sink. One may wonder whether the additional state that this process involves interferes with our minimality proof. The answer is negative: if $\B$ in Theorem~\ref{C is minimal} is not total, then, by Proposition~\ref{there-has-to-be-a-subsafe}, $\A$ has a state $s$ such that $q_{rej} \precsim s$, where $q_{rej}$ is a rejecting sink we need to add to $\B$ if we want to make it total. Thus, $L(\A^s) = \emptyset$, and we may not count the state $s$ if we allow GFG-tNCWs without a total transition function. \hfill \qed}
	\end{rem}

	\subsection{Safe Centralization}\label{sec sc}
	Consider a nice GFG-tNCW $\A = \langle \Sigma, Q_{\A}, q^0_{\A}, \delta_{\A}, \alpha_{\A}\rangle$.
	Recall that $\A$ is safe-centralized if for every two states $q, s\in Q_\A$, if $q \precsim s$, then $q$ and $s$ are in the same safe component.
	In this section we describe how to turn a given nice GFG-tNCW into a nice safe-centralized GFG-tNCW\@. The resulted tNCW is also going to be $\alpha$-homogenous.

	Let $H\subseteq \S(\A)\times \S(\A)$ be such that for all safe components $S, {S}'\in \S(\A)$, we have that $H(S, {S}')$ iff there exist states $q\in S$ and $q' \in {S}'$ such that $q\precsim q'$. That is, when $S \neq {S}'$, then the states $q$ and $q'$ witness that $\A$ is not safe-centralized. Recall that $q\precsim q'$ iff $L(\A^q) = L(\A^{q'})$ and $L_{{\it safe}}(\A^q) \subseteq L_{{\it safe}}(\A^{q'})$. Since language containment for GFG-tNCWs can be checked in polynomial time~\cite{HKR02,KS15}, the first condition can be checked in polynomial time. Since $\A$ is safe deterministic, the second condition reduces to language containment between deterministic automata,
	and can also be checked in polynomial time. Hence, the relation $H$ can be computed in polynomial time.

	\begin{lem}\label{totally-cover-lemma}
		Consider safe components $S, {S}' \in \S(\A)$ such that $H(S, {S}')$. Then, for every $p \in S$ there is $p'\in {S}'$ such that $p \precsim p'$.
	\end{lem}

	\begin{proof}

		Since $H(S, {S}')$, then, by definition, there are states $q\in S$ and $q'\in {S}'$, such that $q \precsim q'$. Let $p$ be a state in $S$. Since $S$ is a safe component, there is a safe run from $q$ to $p$ in $S$. Since $q \precsim q'$, an iterative application of Proposition~\ref{equiv-to-equiv} implies that there is a safe run $r$ from $q'$ to some state $p'$ for which $p \precsim p'$. As $\A$ is normal and $r$ is safe, the run $r$ stays in $S'$; in particular, $p'\in S'$, and we are done.
	\end{proof}

	\begin{lem}\label{H-is-transitive}
		The relation $H$ is transitive: for every safe components $S, {S}', {S}'' \in \S(\A)$, if $H(S, {S}')$ and $H({S}', {S}'')$, then $H(S, {S}'')$.
	\end{lem}

	\begin{proof}
		Let $S, {S}', {S}'' \in \S(\A)$ be safe components of $\A$ such that $H(S, {S}')$ and $H({S}', {S}'')$. Since $H(S, {S}')$, there are states $q\in S$ and $q'\in {S}'$ such that $q \precsim q'$. Now, since $H({S}', {S}'')$, we get by Lemma~\ref{totally-cover-lemma} that for all states in ${S}'$, in particular for $q'$, there is a state $q''\in {S}''$ such that $q' \precsim q''$. The transitivity of $\precsim$ then implies that $q \precsim q''$, and so $H(S, {S}'')$.
	\end{proof}

	We say that a set ${\S}\subseteq \S(\A)$ is a \emph{frontier of $\A$} if for every safe component $S\in \S(\A)$, there is a safe component ${S}'\in {\S}$ with $H(S, {S}')$, and for all safe components $S, {S}' \in {\S}$ such that $S\neq {S}'$, we have  that $\neg H(S, {S}')$ and $\neg H({S}', S)$. Once $H$ is calculated, a frontier of $\A$ can be found in linear time. For example, as $H$ is transitive, we can take one vertex from each ergodic SCC in the graph $\zug{\S(\A),H}$. Note that all frontiers of $\A$ are of the same size, namely the number of ergodic SCCs in this graph.

	Given a frontier $\S$ of $\A$, we define the automaton
	$\B_{\S} = \langle \Sigma, Q_{\S}, q^0_{\S}, \delta_{\S},\alpha_{\S}\rangle$, where $Q_{\S}=\{q\in Q_{\A}: q\in S \text{ for some }S\in {\S} \}$, and the other elements are defined as follows. The initial state $q^0_{\S}$ is chosen such that $q^0_{\S} \sim_\A q^0_\A$. Specifically, if $q^0_\A \in Q_{\S}$, we take $q^0_{\S} = q^0_\A$. Otherwise, by Lemma~\ref{totally-cover-lemma} and the definition of $\S$, there is a state $q' \in Q_{\S}$ such that $q^0_{\A} \precsim q'$, and we take $q^0_{\S}=q'$. The transitions in $\B_{\S}$ are either $\bar{\alpha}$-transitions of $\A$, or $\alpha$-transitions that we add among the safe components in $\S$ in a way that preserves language equivalence. Formally, consider a state $q\in Q_{\S}$ and a letter $\sigma \in \Sigma$. If $\delta^{\bar{\alpha}}_\A(q, \sigma)\neq \emptyset$, then $\delta^{\bar{\alpha}}_\S(q, \sigma) = \delta^{\bar{\alpha}}_\A(q, \sigma)$ and $\delta^{\alpha}_\S(q, \sigma) = \emptyset$.
	If $\delta^{\bar{\alpha}}_\A(q,\sigma) = \emptyset$, then
	$\delta^{\bar{\alpha}}_\S(q, \sigma) = \emptyset$ and $\delta^\alpha_\S(q, \sigma) = \{ q'\in Q_\S: \mbox{ there is } q''\in \delta^\alpha_\A(q, \sigma)  \mbox{ such that } q' \sim_\A q'' \}$.
	Note that $\B_{\S}$ is total and  $\alpha$-homogenous.

	\begin{exa}
		{\rm
			Consider the tDCW $\A$ appearing in Figure~\ref{safe example}. Recall that the dashed transitions are $\alpha$-transitions. Since $\A$ is normal and deterministic, it is nice. By removing the $\alpha$-transitions of $\A$, we get the safe components described in Figure~\ref{A's safe components}.

				\begin{figure}[htb]
		\begin{center}
			\includegraphics[width=.35\textwidth]{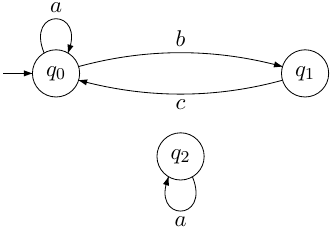}
			\captionof{figure}{The safe components of $\A$.}%
			\label{A's safe components}
		\end{center}
	\end{figure}

			Since $q_2 \precsim q_0$, we have that $\A$ has a single frontier $\S=\{\{q_0,q_1\}\}$. The automaton $\B_{\S}$ appears in Figure~\ref{B figure}. As all the states of $\A$ are equivalent, we direct a $\sigma$-labeled $\alpha$-transition to $q_0$ and to $q_1$, for every state with no $\sigma$-labeled transition in $\S$. \hfill \qed}
			\begin{figure}[htb]
		\begin{center}
			\includegraphics[width=.35\textwidth]{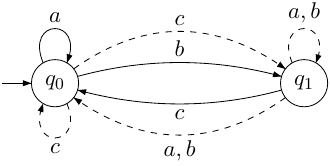}
			\captionof{figure}{The tNCW $\B_\S$, for $\S={\{\{q_0,q_1\}\}}$.}%
			\label{B figure}
		\end{center}
	\end{figure}
\end{exa}

We now prove some properties that will enable us to conclude that a nice GFG-tNCW $\A$ can be turned into a nice, safe-centralized, and $\alpha$-homogenous GFG-tNCW by constructing $\B_{\S}$ for some frontier $\S$ of $\A$. We first extend Proposition~\ref{pruned-corollary} to the setting of $\A$ and $\B_\S$.

	\begin{prop}\label{pruned-corollaryAB}
		Consider states $q$ and $s$ of $\A$ and $\B_{\S}$, respectively, a letter $\sigma \in \Sigma$, and transitions
		$\zug{q, \sigma, q'}$ and $\langle s, \sigma, s'\rangle$ of $\A$ and $\B_{\S}$, respectively.
		If $q \sim_{\A} s$, then $q' \sim_{\A} s'$.
	\end{prop}

	\begin{proof}
		If $\langle s, \sigma, s'\rangle$ is an $\bar{\alpha}$-transition of $\B_{\S}$, then, by the definition of $\Delta_{\S}$, it is also an $\bar{\alpha}$-transition of $\A$. Hence, since $q \sim_{\A} s$ and $\A$ is nice, in particular, semantically deterministic, we get by Proposition~\ref{pruned-corollary} that $q' \sim_{\A} s'$.
		If $\langle s, \sigma, s'\rangle$ is an $\alpha$-transition of $\B_{\S}$, then, by the definition of $\Delta_{\S}$, there is some $s''\in \delta_{\A}(s, \sigma)$ with $s' \sim_{\A} s''$. Again, since $q \sim_{\A} s$ and $\A$ is semantically deterministic, we have by Proposition~\ref{pruned-corollary} that $s'' \sim_{\A} q'$, and thus $s' \sim_{\A} q'$.
	\end{proof}

Note that an iterative application of Proposition~\ref{pruned-corollaryAB} implies that runs of $\A$ and $\B_\S$ on the same finite prefix end in $\A$-equivalent states. Hence, as safe runs in $\B_\S$ exist also in $\A$, one can show that $L(\B_\S)\subseteq L(\A)$.
Conversely, one can show that there is a strategy $g: \Sigma^* \to Q_\S$ such that following $g$ results in accepting all words $w\in L(\A)$ in $\B_\S$.  Essentially, $g$ follows a GFG strategy $f$ for $\A$; that is, whenever $g$ needs to make some nondeterministic choice after reading a prefix $x$, it moves to a state $q'$ with $f(x)\precsim_\A q'$. As $f(w)$ is accepting for all $w\in L(\A)$, then it eventually becomes safe, and thus $g(w)$ cannot make infinitely many bad nondeterministic choices.
The following proposition formalizes the above idea and generalizes it to all pairs of equivalent states.

	\begin{prop}\label{A and B are equivalent}
		Let $q$ and $s$ be states of $\A$ and $\B_{\S}$, respectively, with $q \sim_{\A} s$. It holds that $\B_{\S}^s$ is a GFG-tNCW equivalent to $\A^q$.
	\end{prop}

	\begin{proof}
		We first prove that $L(\B_{\S}^s) \subseteq L(\A^q)$. Consider a word $w=\sigma_1\sigma_2\cdots \in L(\B_{\S}^s)$. Let $s_0,s_1,s_2,\ldots $ be an accepting run of $\B_{\S}^s$ on $w$. Then, there is $i\geq 0$ such that $s_i,s_{i+1},\ldots $ is a safe run of $\B_{\S}^{s_i}$ on the suffix $w[i+1, \infty]$. Let $q_0,q_1,\ldots, q_i$ be a run of $\A^q$ on the prefix $w[1, i]$. Since $q_0 \sim_{\A} s_0$, we get, by an iterative application of Proposition~\ref{pruned-corollaryAB}, that $q_i \sim_{\A} s_i$. In addition, as the run of $\B_{\S}^{s_i}$ on the suffix $w[i+1, \infty]$ is safe, it is also a safe run of $\A^{s_i}$. Hence, $w[i+1, \infty] \in L(\A^{q_i})$, and thus $q_0,q_1,\ldots, q_i$ can be extended to an accepting run of $\A^q$ on $w$.

		Next, we prove that $L(\A^q) \subseteq L(\B_{\S}^s)$ and that $\B_{\S}^s$ is a GFG-tNCW\@. We do this by defining a strategy $g:\Sigma^* \to Q_{\S}$ such that for all words $w \in L(\A^q)$, we have that $g(w)$ is an accepting run of $\B_{\S}^s$ on $w$.
		First, $g(\epsilon)=s$. Then, for $u \in \Sigma^*$ and $\sigma \in \Sigma$, we define $g(u \cdot \sigma)$ as follows. Recall that $\A$ is nice. So, in particular, $\A^q$ is GFG\@. Let $f$ be a strategy witnessing $\A^q$'s GFGness. If $\delta^{\bar{\alpha}}_\S(g(u),\sigma) \neq \emptyset$, then $g(u \cdot \sigma)=q'$ for some $q' \in \delta^{\bar{\alpha}}_\S(g(u),\sigma)$.
		If $\delta^{\bar{\alpha}}_\S(g(u),\sigma) = \emptyset$, then $g(u \cdot \sigma)=q'$ for some state $q' \in Q_{\S}$ such that $f(u \cdot \sigma) \precsim_\A q'$. Note that since $\S$ is a frontier, such a state $q'$ exists. We prove that $g$ is consistent with $\Delta_{\S}$. In fact, we prove a stronger claim, namely for all $u \in \Sigma^*$ and $\sigma \in \Sigma$, we have that $f(u)\sim_\A g(u)$ and $\zug{g(u),\sigma,g(u \cdot \sigma)} \in \Delta_{\S}$.

		The proof proceeds by an induction on $|u|$.
		For the induction base, as $f(\epsilon)=q$, $g(\epsilon)=s$, and $q \sim_\A s$, we are done.
		Given $u$ and $\sigma$, consider a transition $\zug{g(u), \sigma, s'} \in \Delta_{\S}$. Since $\B_{\S}$ is total, such a transition exists.
		We distinguish between two cases. If $\delta^{\bar{\alpha}}_\S(g(u),\sigma) \neq \emptyset$, then, as
		$\B_{\S}$ is $\alpha$-homogenous and safe deterministic, the state $s'$ is the only state in $\delta^{\bar{\alpha}}_\S(g(u),\sigma)$. Hence, by the definition of $g$, we have that $g(u \cdot \sigma)=s'$ and so $\zug{g(u),\sigma,g(u \cdot \sigma)} \in \Delta_{\S}$.
		If $\delta^{\bar{\alpha}}_\S(g(u),\sigma) = \emptyset$, we claim that $g(u \cdot \sigma)  \sim_\A s'$.
		Then, as $s' \in \delta^\alpha_\S(g(u), \sigma)$, the definition of $\Delta_{\S}$ for the case $\delta^{\bar{\alpha}}_\S(g(u),\sigma) = \emptyset$ implies that $\zug{g(u),\sigma,g(u \cdot \sigma)} \in \Delta_{\S}$.
		By the induction hypothesis, we have that $f(u)\sim_\A g(u)$. Hence, as $\zug{f(u),\sigma,f(u \cdot \sigma)} \in \Delta_\A$ and
		$\zug{g(u), \sigma, s'} \in \Delta_{\S}$, we have, by Proposition~\ref{pruned-corollaryAB}, that $f(u \cdot \sigma)\sim_\A s'$.
		Recall that $g$ is defined so that $f(u \cdot \sigma) \precsim_\A g(u \cdot \sigma)$. In particular, $f(u \cdot \sigma) \sim_\A g(u \cdot \sigma)$. Hence, by transitivity of $\sim_\A$, we have that $g(u \cdot \sigma)  \sim_\A s'$, and we are done.
		In addition, by the induction hypothesis, we have that $f(u)\sim_\A g(u)$, and so, in both cases, Proposition~\ref{pruned-corollaryAB} implies that  $f(u \cdot \sigma) \sim_{\A} g(u \cdot \sigma)$.

		It is left to prove that for every infinite word $w = \sigma_1\sigma_2\cdots \in \Sigma^\omega$, if $w\in L(\A^q)$, then $g(w)$ is accepting. Assume that $w\in L(\A^q)$ and consider the run $f(w)$ of $\A^q$ on $w$. Since $f(w)$ is accepting, there is $i\geq 0$ such that $f(w[1, i]),f(w[1, i+1])\ldots $ is a safe run of $\A^{f(w[1,i])}$ on the suffix $w[i+1, \infty]$. We prove that $g(w)$ may traverse at most one $\alpha$-transition when it  reads the suffix $w[i+1, \infty]$. Assume that there is some $j\geq i$ such that $\langle g(w[1, j]), \sigma_{j+1}, g(w[1, j+1])\rangle \in \alpha_\S$. Then, by $g$'s definition, we have that $f(w[1, j+1]) \precsim_\A g(w[1, j+1])$.  Therefore, as $\B_{\S}$ follows the safe components in $\S$,  we have that $L_{{\it safe}}(\A^{f(w[1, j+1])}) \subseteq L_{{\it safe}}(\A^{g(w[1, j+1])}) = L_{{\it safe}}(\B_{\S}^{g(w[1, j+1])})$, and thus $w[j+2, \infty] \in L_{{\it safe}}(\B_{\S}^{g(w[1, j+1])})$. Since $\B_{\S}$ is $\alpha$-homogenous and safe deterministic, there is a single run of $\B^{g(w[1, j+1])}_{\S}$ on $w[j+2, \infty]$, and this is the run that $g$ follows. Therefore, $g(w[1, j+1]),g(w[1, j+2]),\ldots $ is a safe run, and we are done.
	\end{proof}

\begin{rem}\label{all are reachable}
{\rm As we argue below, for every frontier $\S$, all the states of the tNCW $\B_\S$ are reachable.
Since, however, states that are not reachable are easy to detect and their removal does not affect $\B_{\S}$'s language, its other properties mentioned in Proposition~\ref{n sc h}, and arguments  that refer to its construction, we can simply assume that all the states of $\B_\S$ are reachable.

To see that indeed all the states of the tNCW $\B_\S$ are reachable, assume by way of contradiction that there is an unreachable state $q$ of $\B_\S$. Let $P$ be the nonempty set of states in $\B_\S$ that are reachable in $\B_\S$ and are $\A$-equivalent to $q$. In order to reach a contradiction, one can define, using the fact that $\S$ is a frontier, a finite word $z\in \Sigma^*$ such that there is a safe run from $q$ back to $q$ on the input word  $z$ and for all $p\in P$, all runs of $p$ on $z$ are not safe. If such a word $z$ exists, then one can show that $L(\A)\setminus L(\B_\S)$ is nonempty, contradicting their equivalence. \hfill \qed}
	\end{rem}

	The following proposition shows that the modifications applied to $\A$ in order to obtain $\B_{\S}$ maintain the useful properties of $\A$.

	\begin{prop}%
		\label{n sc h}
		For every frontier $\S$, the GFG-tNCW $\B_{\S}$ is nice, safe-centralized, and $\alpha$-homogenous.
	\end{prop}

	\begin{proof}
		It is easy to see that the fact $\A$ is nice implies that $\B_{\S}$ is normal and safe deterministic. %
		Also, as we explained in Remark~\ref{all are reachable}, all the states in $\B_\S$ are reachable.
		Finally, Proposition~\ref{A and B are equivalent} implies that all its states are GFG\@. To conclude that $\B_{\S}$ is nice, we prove below that it is semantically deterministic.  Consider transitions $\langle q, \sigma, s_1\rangle$ and $\langle q, \sigma, s_2\rangle$ in $\Delta_{\S}$. We need to show that $s_1 \sim_{\B_{\S}} s_2$. By the definition of $\Delta_{\S}$, there are transitions $\langle q, \sigma, s'_1\rangle$ and $\langle q, \sigma, s'_2\rangle$ in $\Delta_\A$ for states $s'_1$ and $s'_2$ such that $s_1 \sim_\A s'_1$ and $s_2 \sim_\A s'_2$. As $\A$ is semantically deterministic, we have that $s'_1 \sim_\A s'_2$, thus by transitivity of $\sim_\A$, we get that $s_1 \sim_\A s_2$. Then, Proposition~\ref{A and B are equivalent} implies that $L(\A^{s_1}) = L(\B^{s_1}_{\S})$ and $L(\A^{s_2}) = L(\B^{s_2}_{\S})$, and so we get that $s_1 \sim_{\B_{\S}} s_2$. Thus, $\B_{\S}$ is semantically deterministic.

		As we noted in the definition of its transitions, $\B_{\S}$ is $\alpha$-homogenous.
		It is thus left to prove that $\B_{\S}$ is safe-centralized. Let $q$ and $s$ be states of $\B_{\S}$ such that $q \precsim_{\B_{\S}} s$; that is, $L(\B^q_{\S}) = L(\B^s_{\S})$ and $L_{{\it safe}}(\B^q_{\S})\subseteq L_{{\it safe}}(\B^s_{\S})$. Let $S,T\in \S$ be the safe components of $q$ and $s$, respectively. We need to show that $S = T$. By Proposition~\ref{A and B are equivalent}, we have that $L(\A^q) = L(\B^q_{\S})$ and $L(\A^s) = L(\B^s_{\S})$. As $\B_{\S}$ follows the safe components in ${\S}$, we have that $L_{{\it safe}}(\A^q) = L_{{\it safe}}(\B^q_{\S})$ and $L_{{\it safe}}(\A^s) = L_{{\it safe}}(\B^s_{\S})$. Hence, $q \precsim_{\A} s$, implying $H(S,T)$. Since $\S$ is a frontier, this is possible only when~$S =T$, and we are done.
	\end{proof}

	\begin{thm}
		Every nice GFG-tNCW can be turned in polynomial time into an equivalent nice, safe-centralized, and $\alpha$-homogenous GFG-tNCW\@.
	\end{thm}

	\subsection{Safe-Minimization}\label{sec sd and sm}

	In the setting of finite words, a \emph{quotient automaton} is obtained by merging equivalent states, and is guaranteed to be minimal. In the setting of co-B\"uchi automata, it may not be possible to define an equivalent language on top of the quotient automaton. For example, all the states in the GFG-tNCW $\A$ in Figure~\ref{safe example} are equivalent, and still it is impossible to define its language on top of a single-state tNCW\@. In this section we show that when we start with a nice, safe-centralized, and $\alpha$-homogenous GFG-tNCW $\B$, the transition to a quotient automaton, namely merging of strongly-equivalent states, is well defined and results in a GFG-tNCW equivalent to $\B$ that attains all the helpful properties of $\B$, and is also safe-minimal\footnote{In fact, $\alpha$-homogeneity is not required, but as the GFG-tNCW $\B_\S$ obtained in Section~\ref{sec sc} is $\alpha$-homogenous, which simplifies the proof, we are going to rely on it.}. Then, by Theorem~\ref{C is minimal}, the obtained GFG-tNCW is also minimal.

	Consider a nice, safe-centralized, and $\alpha$-homogenous GFG-tNCW $\B=\zug{\Sigma,Q,q_0,\delta,\alpha}$.
	For a state $q \in Q$, define $[q] = \{ q'\in Q: q \approx_{\B} q' \}$. We define the tNCW $\C = \langle \Sigma, Q_\C, [q_0], \delta_{\C}, \alpha_{\C}\rangle$ as follows. First, $Q_\C=\{[q]: q\in Q\}$. Then, the transition function is such that $\zug{ [q], \sigma, [p]} \in \Delta_\C$ iff there are $q'\in [q]$ and $p'\in [p]$ such that $\langle q', \sigma, p'\rangle \in \Delta$, and $\langle [q], \sigma, [p]\rangle \in \alpha_\C$ iff $\langle q', \sigma, p'\rangle \in \alpha$. Note that $\B$ being $\alpha$-homogenous implies that $\alpha_\C$ is well defined; that is, independent of the choice of $q'$ and $p'$. %
	To see why,  assume that $\langle q', \sigma, p'\rangle \in \bar{\alpha}$, and let $q''$ be a state in $[q]$. As $q' \approx_\B q''$, we have, by Proposition~\ref{equiv-to-equiv}, that there is $p''\in [p]$ such that $\zug{q'', \sigma, p''}\in \bar{\alpha}$. Thus, as $\B$ is $\alpha$-homogenous,  there is no $\sigma$-labeled $\alpha$-transition from $q''$ in $\B$. In particular, there is no $\sigma$-labeled $\alpha$-transition from $q''$  to a state in $[p]$. Note that, by the above, the tNCW $\C$ is $\alpha$-homogenous. Also, if $\langle [q], \sigma, [p]\rangle$ is an $\bar{\alpha}$-transition of $\C$, then for every $q'\in [q]$, there is $p'\in [p]$ such that $\langle q', \sigma, p'\rangle$ is an $\bar{\alpha}$-transition of $\B$. The latter implies by induction the $\supseteq$-direction in the following proposition, stating that a safe run in $\C$ induces a safe run in $\B$. The $\subseteq$-direction follows immediately from the definition of $\C$.

	\begin{prop}\label{same safe obsv}
		For every $[p]\in Q_{\C}$ and every $s\in [p]$, it holds that $L_{{\it safe}}(\B^s) = L_{{\it safe}}(\C^{[p]})$.
	\end{prop}

	We extend Propositions~\ref{pruned-corollary} and~\ref{pruned-corollaryAB} to the setting of $\B$ and $\C$:

	\begin{prop}\label{pruned-corollaryBC}
		Consider states $s \in Q$ and $[p] \in Q_\C$, a letter $\sigma \in \Sigma$, and transitions
		$\zug{s, \sigma, s'}$ and $\langle [p], \sigma, [p']\rangle$ of $\B$ and $\C$, respectively.
		If $s \sim p$, then $s' \sim p'$.
	\end{prop}

	\begin{proof}
		As $\langle [p], \sigma, [p']\rangle$ is a transition of $\C$, there are states $t \in [p]$ and $t'\in [p']$, such that $\langle t, \sigma, t'\rangle \in \Delta$. If $s \sim p$, then $s \sim t$. Since $\B$ is nice, in particular, semantically deterministic, and $\zug{s, \sigma, s'}\in \Delta$, we get by Proposition~\ref{pruned-corollary} that $s' \sim t'$. Thus, as $t' \sim p'$, we are done.
	\end{proof}

		It is not hard to see that $L(\B)\subseteq L(\C)$. Indeed, an accepting run $q_0,q_1,q_2,\ldots$ in $\B$ induces the run $[q_0],[q_1],[q_2],\ldots$ in $\C$. The harder direction is to show that $L(\C)\subseteq L(\B)$. Note that an iterative application of Proposition~\ref{pruned-corollaryBC} implies that runs of $\B$ and $\C$ on the same finite prefix end in states $s\in Q$ and $[p]\in Q_\C$ with $s\sim p$. Hence, as safe runs in $\C$ induce safe runs in $\B$, one can show that $L(\C)\subseteq L(\B)$. Formally, we have the following.

	\begin{prop}\label{final equivalent}
		For every $[p]\in Q_{\C}$ and $s\in [p]$, we have that $\C^{[p]}$ is a GFG-tNCW equivalent to~$\B^s$.
	\end{prop}

	\begin{proof}
		We first prove that $L(\C^{[p]}) \subseteq L(\B^s)$. Consider a word $w=\sigma_1\sigma_2\cdots \in L(\C^{[p]})$. Let $[p_0],[p_1],[p_2],\ldots $ be an accepting run of $\C^{[p]}$ on $w$. Then, there is $i\geq 0$ such that $[p_i],[p_{i+1}],\ldots $ is a safe run of $\C^{[p_i]}$ on the suffix $w[i+1, \infty]$. Let $s_0,s_1,\ldots, s_i$ be a run of $\B^s$ on the prefix $w[1, i]$. Note that $s_0=s$. Since $s_0 \in [p_0]$, we have that $s_0 \sim p_0$, and thus an iterative application of Proposition~\ref{pruned-corollaryBC} implies that $s_i \sim p_i$. In addition, as $w[i+1, \infty]$ is in $L_{{\it safe}}(\C^{[p_i]})$, we get, by Proposition~\ref{same safe obsv}, that $w[i+1, \infty]\in L_{{\it safe}}(\B^{p_i})$. Since $L_{{\it safe}}(\B^{p_i})\subseteq L(\B^{p_i})$ and $s_i \sim p_i$, we have that $w[i+1, \infty]\in L(\B^{s_i})$. Hence, $s_0,s_1,\ldots, s_i$ can be extended to an accepting run of $\B^s$ on $w$.

		Next, we prove that $L(\B^s) \subseteq L(\C^{[p]})$ and that $\C^{[p]}$ is a GFG-tNCW\@. We do this by defining a strategy $h:\Sigma^* \to Q_\C$ such that for all words $w\in L(\B^s)$, we have that $h(w)$ is an accepting run of $\C^{[p]}$ on $w$. We define $h$ as follows. Recall that $\B$ is nice. So, in particular, $\B^{s}$ is GFG\@. Let $g$ be a strategy witnessing $\B^s$'s GFGness. We define $h(u) = [g(u)]$, for every finite word $u\in \Sigma^*$. Consider a word $w \in L(\B^s)$, and consider the accepting run $g(w) = g(w[1, 0]),g(w[1, 1]),g(w[1, 2]),\ldots$ of $\B^s$ on $w$. Note that by the definition of $\C$, we have that $h(w) = [g(w[1, 0])], [g(w[1, 1])], [g(w[1, 2])],\ldots$ is an accepting run of $\C^{[p]}$ on $w$, and so we are done.
	\end{proof}

		The following proposition shows that the transition from $\B$ to $\C$ maintains the useful properties of $\B$. 	The considerations are similar to these applied
		in the proof of Proposition~\ref{n sc h}. In particular, for safe-minimality, note that for states $q$ and $s$
		of $\B$, we have that $[q]\approx [s]$ iff $[q]\preceq [s]$ and $[s]\preceq [q]$. Thus, it is sufficient to prove that if
		$[q]\preceq [s]$, then $q\preceq s$.

	\begin{prop}%
		\label{n sc sm}
		The GFG-tNCW $\C$ is nice, safe-centralized, safe-minimal, and $\alpha$-homogenous.
	\end{prop}

	\begin{proof}
		To begin with, as $\B$ is nice, it is easy to see that all the states in $\C$ are reachable. In addition, Proposition~\ref{final equivalent} implies that all its states are GFG\@. We show next that $\C$ is safe deterministic. Let $\zug{[q], \sigma, [s_1]}$ and $\zug{[q], \sigma, [s_2]}$ be $\bar{\alpha}$-transitions of $\C$, and let $q', q''\in [q], s'_1\in [s_1], s'_2\in [s_2]$ be such that $t_1 = \zug{q', \sigma, s'_1}$ and $t_2 = \zug{q'', \sigma, s'_2}$ are $\bar{\alpha}$-transitions of $\B$. Now, as $q' \approx q''$ and $\B$ is nice, in particular, safe deterministic, Proposition~\ref{equiv-to-equiv} implies that $s'_1 \approx s'_2$, and thus $\C$ is safe deterministic. Next, we prove that $\C$ is normal. Consider a safe run $[p_0],[p_1],\ldots, [p_n]$ of $\C^{[p_0]}$ on the finite word $w = \sigma_1\sigma_2\cdots \sigma_n\in \Sigma^*$. We need to show that there is a safe run from $[p_n]$ to $[p_0]$ in $\C$. As mentioned in $\C$'s definition, a safe run in $\C$ induces a safe run in $\B$. Thus, there is a safe run $s_0,s_1, \ldots, s_n$ of $\B^{s_0} $ on $w$, where $s_0\in [p_0]$ and $s_n\in [p_n]$. Now as $\B$ is nice, in particular, normal, we have that there is a safe run $r_0, r_1, \ldots, r_m$ of $\B^{r_0}$ on $w'=\sigma'_1\sigma'_2\cdots \sigma'_m$, where $r_0 = s_n$ and $r_m = s_0$, and since $s_n\in [p_n]$ and $s_0\in [p_0]$, we get by $\C$'s definition that $[r_0],[r_1],\ldots, [r_m]$ is a safe run on $w'$ from $[r_0] = [s_n] = [p_n]$ to $[r_m] = [s_0] = [p_0]$, and so $\C$ is normal.
		To conclude that $\C$ is nice, we prove below that it is semantically deterministic. Consider the transitions $\langle [q], \sigma, [s_1]\rangle$ and $\langle [q], \sigma, [s_2] \rangle$ in $\C$. We need to show that $L(\C^{[s_1]}) = L(\C^{[s_2]})$. By $\C$'s definition, there are states $q', q''\in [q]$, $s'_1\in [s_1]$ and $s'_2\in [s_2]$ with the transitions $\langle q', \sigma, s'_1\rangle$ and $\langle q'', \sigma, s'_2\rangle$ in $\B$. Now as $\B$ is semantically deterministic and $L(\B^{q'}) = L(\B^{q''})$, we have by Proposition~\ref{pruned-corollary} that $L(\B^{s'_1}) = L(\B^{s'_2})$, and thus Proposition~\ref{final equivalent} implies that $L(\C^{[s_1]}) = L(\B^{s'_1}) = L(\B^{s'_2}) = L(\C^{[s_2]})$, and so $\C$ is semantically deterministic.

		To prove that $\C$ is safe-minimal, note that $[q]\approx [s]$ iff $[q]\precsim [s]$ and $[s]\precsim [q]$. Thus, it is sufficient to prove that if $[q] \precsim [s]$ then $q \precsim s$. If $L(\C^{[q]}) = L(\C^{[s]})$ and $L_{{\it safe}}(\C^{[q]}) \subseteq L_{{\it safe}}(\C^{[s]})$, then by Proposition~\ref{final equivalent}, we have that $L(\B^q) = L(\C^{[q]}) = L(\C^{[s]}) = L(\B^s)$, and thus $L(\B^q) = L(\B^s)$. Similarly, by Proposition~\ref{same safe obsv}, we have that $L_{{\it safe}}(\B^q) = L_{{\it safe}}(\C^{[q]}) \subseteq L_{{\it safe}}(\C^{[s]}) = L_{{\it safe}}(\B^s)$, and thus $q \precsim s$.

		As we noted in the definition of its transitions, $\C$ is $\alpha$-homogenous.
		It is thus left to prove that $\C$ is safe-centralized. If $[q] \precsim [s]$, then as seen previously, $q \precsim s$, and since $\B$ is safe-centralized, we get that $q$ and $s$ are in the same safe component of $\B$. Therefore, there is a safe run, $r_0,r_1,\ldots, r_n$, from $q$ to $s$ in $\B$, and thus we have by $\C$'s definition that $[r_0],[r_1],\ldots, [r_n]$ is a safe run from $[q]$ to $[s]$ in $\C$, implying that $[q]$ and $[s]$ are in the same safe component as $\C$ is normal, and so we are done.
	\end{proof}

	We can thus conclude with the following.

	\begin{thm}
		Every nice, safe-centralized, and $\alpha$-homogenous GFG-tNCW can be turned in polynomial time into an equivalent nice, safe-centralized,  safe-minimal, and $\alpha$-homogenous GFG-tNCW\@.
	\end{thm}

	\begin{exa}
		{\rm The safe languages of the states $q_0$ and $q_1$ of the GFG-tNCW $\B_\S$ from Figure~\ref{B figure} are different. Thus, $q_0 \not \approx q_1$, and applying safe-minimization to $\B_\S$ results in the GFG-tNCW $\C$ identical to $\B_\S$. \hfill \qed}
	\end{exa}

	\section{Canonicity in GFG-tNCWs}%
\begin{figure}[htb]
		\begin{center}
			\includegraphics[width=.8\textwidth]{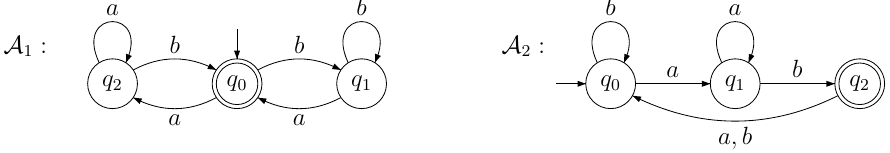}
			\caption{The DCWs $\A_1$ and $\A_2$.}%
			\label{2min dcws}
		\end{center}
	\end{figure}

	\label{sec canonical rep}
	In this section we study canonicity for GFG-tNCWs.
	Recall that unlike the case of finite words, for $\omega$-regular languages, no canonical minimal deterministic automaton exists~\cite{Kup15}.
	This classical result refers to automata with state-based acceptance condition. There, for an automaton $\A=\zug{\Sigma,Q,q_0,\delta,\alpha}$, we have that $\alpha \subseteq Q$. In \emph{B\"uchi} automata (NBWs or DBWs), a run $r$ is accepting iff ${\it inf}(r)\cap \alpha \neq \emptyset$, where now ${\it inf}(r)$ is the set of states that $r$ visits infinitely often. Thus, $r$ is accepting if it visits states in $\alpha$ infinitely often. Dually, in \emph{co-B\"uchi} automata (NCWs or DCWs), a run $r$ is accepting iff ${\it inf}(r)\cap \alpha = \emptyset$.
	No canonicity means that an $\omega$-regular language may have different minimal DCWs or DBWs.
	Consider for example the DCWs $\A_1$ and $\A_2$ appearing in Figure~\ref{2min dcws}. Both are minimal DCWs for the language $L=(a+b)^* \cdot (a^\omega + b^\omega)$ (``only finitely many $a$'s or only finitely many $b$'s''; it is easier to see this by considering the dual DBWs, for ``infinitely many $a$'s and infinitely many $b$'s'').

	Since all the states of $\A_1$ and $\A_2$ recognize the language $L$ and may serve as initial states, Figure~\ref{2min dcws} actually presents six different DCWs for $L$, and more three-state DCWs for $L$ exist. The DCWs $\A_1$ and $\A_2$, are however ``more different'' than variants of $\A_1$ obtained by changing the initial state: they have a different structure, or more formally, there is no isomorphism between their graphs.

	We are going to show that unlike the case of DCWs and DBWs, GFG-tNCWs do have a canonical form.
	Our positive results about canonicity of GFG-tNCWs are related to the miniminization algorithm, and
	we first show that the sufficient conditions for minimality of nice GFG-tNCWs specified in
	Theorem~\ref{C is minimal} are necessary.

	\begin{thm}%
		\label{minimal automaton has all properties}
		Nice minimal GFG-tNCWs are safe-centralized and safe-minimal.
	\end{thm}

	\begin{proof}
		Consider a nice minimal GFG-tNCW $\A$. We argue that if $\A$ is not safe-centralized or not safe-minimal, then it can be minimized further by the minimization algorithm described in Section~\ref{sc minimization}.
		Assume first that $\A$ is not safe-centralized. Then, there are two different safe components $S, S' \in \S(\A)$ and states $q \in S$ and $q' \in S'$ such that $q \precsim q'$. Then, $H(S, S')$, implying that the safe components in $\S(\A)$ are not a frontier. Then, safe-centralization, as described in Section~\ref{sec sc}, minimizes $\A$ further. Indeed, in the transition to the automaton $\B_\S$, at least one safe component in $\S(\A)$ is removed from $\A$ when the frontier $\S$ is computed.
		Assume now that $\A$ is safe-centralized. Then, for every two different safe components $S, S' \in \S(\A)$, it holds that $\neg H(S, S')$ and $\neg H(S', S)$. Hence, every strict subset of $\S(\A)$ is not a frontier. Thus, $\S(\A)$ is the only frontier of $\A$. Hence,  the automaton $\B_\S$, constructed in Section~\ref{sec sc}, has $\S=\S(\A)$, and is obtained from $\A$ by adding $\alpha$-transitions that do not change the languages and safe languages of its states.
		Accordingly, $\B_\S$ is safe-minimal iff $\A$ is safe-minimal. Therefore, if $\A$ is not safe-minimal, then  safe-minimizing $\B_\S$, as in Section~\ref{sec sd and sm}, merges at least two different states. Hence, also in this case, $\A$ is minimized further.
	\end{proof}

	We formalize relations between tNCWs by means of \emph{isomorphism} and \emph{safe isomorphism}.
	Consider two tNCWs $\A=\zug{\Sigma,Q_\A,q^0_\A,\delta_\A,\alpha_\A}$ and $\B=\zug{\Sigma,Q_\B,q^0_\B,\delta_\B,\alpha_\B}$, and a bijection $\kappa: Q_\A \to Q_\B$. We say that $\kappa$ is:
	\begin{itemize}[label=$\triangleright$]

		\item
		\emph{$\alpha$-transition respecting}, if $\kappa$ induces a bijection between the $\alpha$-transitions of $\A$ and $\B$. Formally, for all states $q,q' \in Q_\A$ and letter $\sigma \in \Sigma$, we have that  $q' \in \delta^{\alpha_\A}_\A(q, \sigma)$ iff $\kappa(q') \in \delta^{\alpha_\B}_\B(\kappa( q), \sigma)$.

		\item
		\emph{$\bar{\alpha}$-transition respecting}, if $\kappa$ induces a bijection between the $\bar{\alpha}$-transitions of $\A$ and $\B$. Formally, for all states $q,q' \in Q_\A$ and letter $\sigma \in \Sigma$, we have that  $q' \in \delta^{\bar{\alpha}_\A}_\A(q, \sigma)$ iff $\kappa(q') \in \delta^{\bar{\alpha}_\B}_\B(\kappa( q), \sigma)$.

	\end{itemize}

    \noindent
	Then, $\A$ and $\B$ are \emph{safe isomorphic} if there is a bijection $\kappa: Q_\A \to Q_\B$ that is $\bar{\alpha}$-transition respecting. If, in addition, $\kappa$ is $\alpha$-transition respecting, then $\A$ and $\B$ are \emph{isomorphic}. Note that if $\kappa$ is $\bar{\alpha}$-transition respecting, then for every state $q \in Q_\A$, it holds that
	$q$ and $\kappa(q)$ have equivalent safe languages. 
	Also, if $\kappa$ is both $\alpha$-transition respecting and $\bar{\alpha}$-transition respecting, then for every state $q \in Q_\A$, we have that
	$q \approx \kappa(q)$.

	\subsection{Safe isomorphism}

	Theorem~\ref{minimal automaton has all properties} suggests that safe-minimization and safe-centralization characterize nice minimal GFG-tNCWs. Hence, if we take two nice minimal equivalent GFG-tNCWs $\A$ and $\B$, and apply Lemma~\ref{injection} to both of them, one can conclude that the injection $\eta: \S(\A)\to \S(\B)$ is actually a bijection. The following theorem uses the latter fact in order to show that $\eta$ induces a bijection $\kappa: Q_\A \to Q_\B$ that is $\bar{\alpha}$-transition respecting.

	\begin{thm}%
		\label{safe isomorphic}
		Every two equivalent, nice, and minimal GFG-tNCWs are safe isomorphic.
	\end{thm}

	\begin{proof}
		Consider two equivalent, nice, and minimal GFG-tNCWs $\A$ and $\B$. By Theorem~\ref{minimal automaton has all properties}, $\A$ is safe-minimal and safe-centralized. Hence, by Lemma~\ref{injection}, there is an injection $\eta: \S(\A) \to \S(\B)$ such that for every safe component $T\in \S(\A)$, it holds that $|T|\leq |\eta(T)|$. For a safe component $T\in \S(\A)$, let $p_T$ be some state in $T$. By Proposition~\ref{there is the same safe}, there are states $q_T \in Q_{\A}$ and $s_T \in Q_{\B}$ such that $p_T \precsim q_T$ and $q_T \approx s_T$. Since $\A$ is safe-centralized, the state $q_T$ is in $T$, and in the proof of Lemma~\ref{injection}, we defined $\eta(T)$ to be the safe component of $s_T$ in $\B$. Likewise, $\B$ is safe-minimal and safe-centralized, and there is an injection $\eta': \S(\B) \to \S(\A)$. %
		The existence of the two injections implies that $|\S(\A)| = |\S(\B)|$. Thus, the injection $\eta$ is actually a bijection. Hence,

		\[|\A| = \sum\limits_{T\in \S(\A)} |T| \leq \sum\limits_{T\in \S(\A)} |\eta(T)| = \sum\limits_{T' \in \S(\B)} |T'| =  |\B|\]

		Indeed, the first inequality follows from the fact $|T| \leq |\eta(T)|$, and the second equality follows from the fact that $\eta$ is a bijection.
		Now, as $\A$ and $\B$ are both minimal, we have that $|\A|=|\B|$, and so it follows that for every safe component $T\in \S(\A)$, we have that $|T| = |\eta(T)|$.
		We use the latter fact in order to show that $\eta$ induces a bijection $\kappa: Q_\A \to Q_\B$ that is $\bar{\alpha}$-transition respecting.

		Consider a safe component $T\in \S(\A)$.
		By Lemma~\ref{injection}, we have that $|T| \leq |\eta(T)|$. The proof of the lemma associates with a safe run  $r_T = q_0, q_1, \ldots, q_m$ of $\A$ that traverses all the states in the safe component $T$, a safe run $r_{\eta(T)} = s_0, s_1, \ldots, s_m$ of $\B$ that traverses states in $\eta(T)$ and $q_i \approx s_i$, for every $0\leq i\leq m$. Moreover, if $0 \leq i_1,i_2 \leq m$ are such that $q_{i_1} \not \approx q_{i_2}$, then $s_{i_1} \not \approx s_{i_2}$.
		Let $\kappa_T: T \to \eta(T)$ be a function that maps each state $q_i$ in $r_T$ to the state $s_i$ in $r_{\eta(T)}$.  If there are $0\leq i_1, i_2\leq m$ with $q_{i_1} = q_{i_2}$, then $q_{i_1}\approx  q_{i_2}$, and thus by transitivity of $\approx$, it holds that $s_{i_1}\approx  s_{i_2}$. Then, safe-minimality of $\B$ implies that $s_{i_1} = s_{i_2}$, and thus $\kappa_T$ is a well defined function from $T$ to $\eta(T)$.
		Since $\A$ is safe-minimal, every two states in $T$ are not strongly equivalent. Therefore, the function $\kappa_T$  is an injection from $T$ to $\eta(T)$. Thus, as $|T| = |\eta(T)|$, the injection $\kappa_T$ is actually a bijection.
		The desired bijection $\kappa$ is then the union of the bijections $\kappa_T$ for $T\in \S(\A)$.

		Clearly, as $\eta: \S(\A) \to \S(\B)$ is a bijection, the function $\kappa$ is a bijection from $Q_\A$ to $Q_\B$. We prove that $\kappa$ is $\bar{\alpha}$-transition respecting.
		Consider states $q,q' \in Q_\A$ and a letter $\sigma \in \Sigma$ such that $\zug{q, \sigma, q'}$ is an $\bar{\alpha}$-transition of $\A$.
		Let $T$ be $q$'s safe component. By the definition of $\kappa_T$, we have that $q \approx \kappa_T(q)$. By Proposition~\ref{equiv-to-equiv}, there is an $\bar{\alpha}$-transition of $\B$ of the form $t = \zug{\kappa(q), \sigma, s'}$, where $q' \approx s'$. As $t$ is an $\bar{\alpha}$-transition of $\B$, we know that $s'$ is in $\eta(T)$. Recall that $\B$ is safe-minimal; in particular, there are no strongly-equivalent states in $\eta(T)$. Hence, $s' = \kappa(q')$, and so $\zug{\kappa(q), \sigma, \kappa(q')}$ is an $\bar{\alpha}$-transition of $\B$. Likewise, if $\zug{\kappa(q), \sigma, \kappa(q')}$ is an $\bar{\alpha}$-transition of $\B$, then $\zug{q, \sigma, q'}$ is an $\bar{\alpha}$-transition of $\A$, and so we are done.
	\end{proof}

	\subsection{Isomorphism}

	Theorem~\ref{safe isomorphic} implies that all nice minimal GFG-tNCWs for a given language are safe isomorphic. We continue and show that it is possible to make these GFG-tNCWs isomorphic.
	We propose two canonical forms that guarantee isomorphism. Both forms are based on saturating the GFG-tNCW with $\alpha$-transitions. One adds as many $\alpha$-transitions as possible, and the second does so in a way that preserves $\alpha$-homogeneity.

	Consider a nice GFG-tNCW $\A=\zug{\Sigma,Q,q_0,\delta,\alpha}$. We say that a triple $\zug{q, \sigma, s} \in Q \times \Sigma \times Q$ is an \emph{allowed transition} in $\A$ if there is a state $s' \in Q$ such that $s \sim s'$ and $\zug{q, \sigma, s'}\in \Delta$.  
	Note that since for all states $s \in Q$, we have that $s \approx s$, then all the transitions in $\Delta$ are allowed.
	We now define two types of $\alpha$-maximality.
	\begin{itemize}[label=$\triangleright$]
		\item
		We say that $\A$ is \emph{$\alpha$-maximal} if all allowed transitions in $Q \times \Sigma \times Q$ are in $\Delta$.
		\item
		We say that $\A$ is \emph{$\alpha$-maximal up to homogeneity} if $\A$ is $\alpha$-homogenous, and for every state $q \in Q$ and letter $\sigma \in \Sigma$, if $q$ has no outgoing $\sigma$-labeled $\bar{\alpha}$-transitions, then all allowed transitions in $\{q\} \times \{\sigma\} \times Q$ are in $\Delta$.
	\end{itemize}

    \noindent
	Thus, $\alpha$-maximal automata include all allowed transitions, and $\alpha$-maximal up to homogeneity automata include all allowed transitions as long as their inclusion does not conflict with $\alpha$-homogeneity.

\begin{figure}[htb]
		\begin{center}
			\includegraphics[width=.8\textwidth]{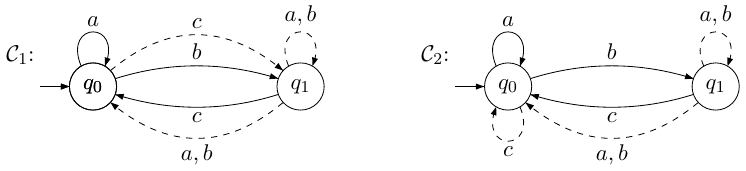}
			\captionof{figure}{Two safe isomorphic yet not isomorphic nice minimal equivalent GFG-tNCWs.}%
			\label{safe iso min}
		\end{center}
	\end{figure}

	\begin{exa}%
		\label{not iso}
		{\rm
			Recall the minimal GFG-tNCW $\B_{\S}$ appearing in Figure~\ref{B figure}.
			The GFG-tNCWs $\C_1$ and $\C_2$ in Figure~\ref{safe iso min} are obtained from $\B_\S$ by removing a $c$-labeled $\alpha$-transition from $q_0$. This does not change the language and results in two nice minimal equivalent GFG-tNCWs that are safe isomorphic yet
			are not $\alpha$-maximal nor $\alpha$-maximal up to homogeneity. \hfill \qed}
	\end{exa}

	We now prove that both types of $\alpha$-maximality guarantee isomorphism.

	\begin{thm}%
		\label{max imp iso}
		Every two equivalent, nice, minimal, and $\alpha$-maximal GFG-tNCWs are isomorphic.
	\end{thm}

	\begin{proof}
		Consider two equivalent, nice, minimal, and $\alpha$-maximal GFG-tNCWs $\C_1$ and $\C_2$. By Theorem~\ref{safe isomorphic}, we have that $\C_1$ and $\C_2$ are safe isomorphic. Thus, there is a bijection $\kappa: Q_{\C_1} \to Q_{\C_2}$ that is $\bar{\alpha}$-transition respecting. The bijection $\kappa$ was defined such that $q \approx \kappa(q)$, for every state $q\in Q_{\C_1}$. We show that $\kappa$ is also $\alpha$-transition respecting. Let $\zug{q, \sigma, s}$ be an $\alpha$-transition of $\C_1$. Then, as $\kappa$ is $\bar{\alpha}$-transition respecting, and $\zug{q, \sigma, s}$ is not an $\bar{\alpha}$-transition in $\C_1$, the triple $\zug{\kappa(q), \sigma, \kappa(s)}$ cannot be an $\bar{\alpha}$-transition in $\C_2$. We show that $\zug{\kappa(q), \sigma, \kappa(s)}$ is a transition in $\C_2$, and thus it has to be an $\alpha$-transition. As $\C_2$ is total, there is a transition $ \zug{\kappa(q), \sigma, s'}$ in $\C_2$. As $q \sim \kappa(q)$ and both automata are nice, in particular, semantically deterministic, Proposition~\ref{pruned-corollary} then implies that $s \sim s'$. Now since $s \sim \kappa(s)$, we get by the transitivity of $\sim$ that $s' \sim \kappa(s)$. Therefore, the existence of the transition $\zug{\kappa(q), \sigma, s'}$ in $\C_2$, implies that the transition $\zug{\kappa(q), \sigma, \kappa(s)} $ is an allowed transition, and so $\alpha$-maximality of $\C_2$ implies that it is also a transition in $\C_2$.  Likewise, if $\zug{\kappa(q), \sigma, \kappa(s)}$ is an $\alpha$-transition in $C_2$, then $\zug{q, \sigma, s}$ is an $\alpha$-transition in $C_1$, and so we are done.
	\end{proof}

	\begin{thm}%
		\label{max hom imp iso}
		Every two equivalent, nice, minimal, and $\alpha$-maximal up to homogeneity GFG-tNCWs are isomorphic.
	\end{thm}

	\begin{proof}
		The proof is identical to that of 	Theorem~\ref{max imp iso}, except that we also have to prove that $\kappa(q)$ has no outgoing $\sigma$-labeled $\bar{\alpha}$-transitions in $\C_2$. To see this, assume by way of contradiction that there is an $\bar{\alpha}$-transition $\zug{\kappa(q), \sigma, s'}$ in $\C_2$. Then, as $q \approx \kappa(q)$, Proposition~\ref{equiv-to-equiv} implies that $q$ has an outgoing $\sigma$-labeled $\bar{\alpha}$-transition in $\C_1$,  contradicting the fact that $\C_1$ is $\alpha$-homogenous.
	\end{proof}

	\subsection{Obtaining Canonical Minimal GFG-tNCWs}%
	\label{sec obtain}

	In this section we show how the two types of canonical minimal GFG-tNCWs can be obtained in polynomial time.
	We start with $\alpha$-maximality up to homogeneity and show that the minimization construction described in Section~\ref{sc minimization} results in GFG-tNCWs that are $\alpha$-maximial up to homogeneity. We continue with $\alpha$-maximality, and show that adding allowed transitions to a GFG-tNCW does not change its language, and conclude that $\alpha$-maximization can be performed on top of the minimization construction described in Section~\ref{sc minimization}.

	\subsubsection{Obtaining canonical minimal \texorpdfstring{$\alpha$-maximal}{alpha-maximal} up to homogeneity GFG-tNCWs}

	\begin{thm}%
		\label{prop canonical form}
		Consider a nice GFG-tNCW $\A$, and let $\C$ be the minimal GFG-tNCW generated from $\A$ by the minimization construction described in Section~\ref{sc minimization}. Then, $\C$ is $\alpha$-maximal up to homogeneity.
	\end{thm}

	\begin{proof}
		Consider the minimization construction described in Section~\ref{sc minimization}. We first show that the safe-centralized GFG-tNCW $\B_{\S}$, defined in Section~\ref{sec sc}, is $\alpha$-maximal up to homogeneity. Then, we show that $\alpha$-maximality up to homogeneity is maintained in the transition to the GFG-tNCW $\C$, defined in Section~\ref{sec sd and sm}.

		By Proposition~\ref{n sc h}, we know that $\B_\S$ is $\alpha$-homogenous. Assume that $q$ is a state in $\B_\S$ with no outgoing $\sigma$-labeled $\bar{\alpha}$-transitions, and assume that $\zug{q, \sigma, s}$ is an allowed transition. We need to show that $\zug{q, \sigma, s}$ is a transition in $\B_\S$.
		As $\zug{q, \sigma, s}$ is an allowed transition, there is a transition $\zug{q, \sigma, s'}$ in $\B_\S$ with $s \sim_{\B_\S} s'$, and by the assumption, $\zug{q, \sigma, s'}$ has to be an $\alpha$-transition.
		By the definition of the transition function of $\B_\S$, we have that $s' \sim_\A q'$ for some state $q' \in \delta^{\alpha}_\A(q, \sigma)$. %
		Now, as $s \sim_{\B_\S} s'$, Proposition~\ref{A and B are equivalent} implies that $L(\A^{s}) = L(\B^{s}_\S) = L(\B^{s'}_\S) = L(\A^{s'})$; that is, $s\sim_\A s'$, and since $s'\sim_\A q'$, we get by the transitivity of $\sim_\A$ that $s \sim_\A q'$, and so $\zug{q, \sigma, s}$ is a transition in $\B_\S$.

		Next, we show that the GFG-tNCW $\C$ is $\alpha$-maximal up to homogeneity. By Proposition~\ref{n sc sm}, we have that $\C$ is $\alpha$-homogenous.
		Assume that $[q]$ is a state in $\C$ with no outgoing $\sigma$-labeled $\bar{\alpha}$-transitions, and assume that $\zug{[q], \sigma, [s]}$ is an allowed transition. We need to show that $\zug{[q], \sigma, [s]}$ is a transition in $\C$.
		As $\zug{[q], \sigma, [s]}$ is an allowed transition, there is a transition $\zug{[q], \sigma, [s']}$ in $\C$ with $[s] \sim [s']$.
		Thus, by Proposition~\ref{final equivalent}, we have that $L(\B^{s}_\S) = L(\C^{[s]}) = L(\C^{[s']}) = L(\B^{s'}_\S)$; that is, $s'\sim_{\B_\S} s$.
		By the assumption, $\zug{[q], \sigma, [s']}$ has to be an $\alpha$-transition. Therefore, by the definition of $\C$, there are states $q''\in [q]$ and $s''\in [s']$, such that $\langle q'', \sigma, s''\rangle$ is an $\alpha$-transition in $\B_\S$. Now, by transitivity of $\sim_{\B_\S}$ and the fact that $s'' \sim_{\B_\S} s'$, we get that $s''\sim_{\B_\S} s$. Finally, as $\B_\S$ is $\alpha$-homogenous, we get that $q''$ has no outgoing $\sigma$-labeled $\bar{\alpha}$-transitions in $\B_\S$, and so by the $\alpha$-maximality up to homogeneity of $\B_\S$, we have that $\langle q'', \sigma, s\rangle$ is a transition in $\B_\S$. Therefore, by the definition of $\C$,  we have that $\langle  [q], \sigma, [s]\rangle$ is a transition in $\C$, and we are done.
	\end{proof}

	We can thus conclude with the following.

	\begin{thm}
		Every GFG-tNCW $\A$ can be canonized into a nice, minimal, and $\alpha$-maximal up to homogeneity GFG-tNCW in polynomial time.
	\end{thm}

	\subsubsection{Obtaining canonical minimal \texorpdfstring{$\alpha$-maximal}{alpha-maximal} GFG-tNCWs}

	Consider a nice GFG-tNCW $\A = \zug{\Sigma, Q, q_0, \delta, \alpha}$.
	Recall that the transition function $\delta$ can be viewed as a transition relation $\Delta \subseteq Q\times \Sigma \times Q$, where for every two states $q,s\in Q$ and letter $\sigma\in \Sigma$, we have that $\langle q, \sigma, s \rangle \in \Delta$ iff $s\in \delta(q, \sigma)$.
	We say that a set of triples  $\E \subseteq \left(Q\times \Sigma \times Q\right) \setminus \Delta$
	is an \emph{allowed set} if all the triples in it are allowed transitions in $\A$.
	For every allowed set  $\E \subseteq \left(Q\times \Sigma \times Q \right)\setminus \Delta$,
	we define the tNCW $\A_\E = \zug{\Sigma, Q_, q_0, \delta_\E, \alpha_\E}$, where $\Delta_\E = \Delta \cup \E$ and $\alpha_\E = \alpha \cup \E$. Clearly, as $\A$ and $\A_\E$ have the same set of states and the same set of $\bar{\alpha}$-transitions, they have equivalent safe languages.

	In Propositions~\ref{C and CE are equivalent} and~\ref{ae is nice} below, we prove that for every allowed set $\E$, we have that $\A_{\E}$ is a nice GFG-tNCW equivalent to $\A$.
	We first extend Proposition~\ref{pruned-corollary} to the setting of $\A$ and $\A_\E$:

	\begin{prop}\label{pruned-corollaryC}
		Consider states $q$ and $s$ of $\A$ and $\A_\E$, respectively, a letter $\sigma \in \Sigma$, and transitions
		$\zug{q, \sigma, q'}$ and $\langle s, \sigma, s'\rangle$ of $\A$ and $\A_\E$, respectively.
		If $q \sim_{\A} s$, then $q' \sim_{\A} s'$.
	\end{prop}

	\begin{proof}
		If $\langle s, \sigma, s'\rangle \notin \E$, then, by the definition of $\Delta_{\E}$, it is also a transition of $\A$. Hence, since $q \sim_{\A} s$ and $\A$ is nice, in particular, semantically deterministic, Proposition~\ref{pruned-corollary} implies  that $q' \sim_{\A} s'$.
		If $\langle s, \sigma, s'\rangle \in \E$, then
		it is an allowed transition of $\A$.  Therefore, there is a state $p'\in Q$ such that $s' \sim_\A p'$ and $\zug{s, \sigma, p'}\in \Delta$. As $q\sim_\A s$ and $\A$ is semantically deterministic, Proposition~\ref{pruned-corollary}  implies that $q'\sim_{\A} p'$. Therefore, using the fact that $p'\sim_\A s'$, the transitivity of $\sim_\A$ implies that $q' \sim_\A s'$, and so we are done.
	\end{proof}

	Now, we can iteratively apply Proposition~\ref{pruned-corollaryC} and show that equivalent states in $\A$ stay equivalent in $\A_\E$, which is GFG\@.

	\begin{prop}\label{C and CE are equivalent}
		Let $p$ and $s$ be states of $\A$ and $\A_\E$, respectively, with $p \sim_{\A} s$. Then, $\A^s_\E$ is a GFG-tNCW equivalent to $\A^p$.
	\end{prop}

	\begin{proof}
		We first prove that $L(\A^s_\E) \subseteq L(\A^p)$. Consider a word $w=\sigma_1\sigma_2\cdots \in L(\A^s_\E)$, and let $s_0,s_1,s_2,\ldots $ be an accepting run of $\A^s_\E$ on $w$. Then, there is $i\geq 0$ such that $s_i,s_{i+1},\ldots $ is a safe run of $\A^{s_i}_\E$ on the suffix $w[i+1, \infty]$. Let $p_0,p_1,\ldots, p_i$ be a run of $\A^p$ on the prefix $w[1, i]$. Since $p_0 \sim_{\A} s_0$, we get, by an iterative application of Proposition~\ref{pruned-corollaryC}, that $p_i \sim_{\A} s_i$. In addition, as the run of $\A^{s_i}_\E$ on the suffix $w[i+1, \infty]$ is safe, it is also a safe run of $\A^{s_i}$. Hence, $w[i+1, \infty] \in L(\A^{p_i})$, and thus $p_0,p_1,\ldots, p_i$ can be extended to an accepting run of $\A^p$ on $w$.

		Next, as $\A$ is nice, all of its states are GFG, in particular, there is a strategy $f^s$ witnessing $\A^s$'s GFGness. Recall that $\A$ is embodied in $\A_\E$. Therefore, every run in $\A$ exists also in $\A_\E$. Thus, as $p \sim_\A s$, we get that for every word $w\in L(\A^p)$, the run $f^s(w)$ is an accepting run of $\A^s$ on $w$, and thus is also an accepting run of $\A^s_\E$ on $w$. Hence, $L(\A^p) \subseteq L(\A^s_\E)$  and $f^s$ witnesses $\A^s_\E$'s GFGness.
	\end{proof}

	\begin{prop}%
		\label{ae is nice}
		For every allowed set $\E$, the GFG-tNCW $\A_\E$ is nice.
	\end{prop}

	\begin{proof}
		It is easy to see that the fact $\A$ is nice implies that $\A_\E$ is normal and safe deterministic. Also, as $\A$ is embodied in $\A_\E$ and both automata have the same state-space and initial states, then all the states in $\A_\E$ are reachable.
		Finally, Proposition~\ref{C and CE are equivalent} implies that all the states in $\A_\E$ are GFG\@. To conclude that $\A_\E$ is nice, we prove below that it is semantically deterministic.  Consider transitions $\langle q, \sigma, s_1\rangle$ and $\langle q, \sigma, s_2\rangle$ in $\Delta_{\E}$. We need to show that $s_1 \sim_{\A_\E} s_2$. By the definition of $\Delta_\E$, 	there are transitions $\langle q, \sigma, s'_1\rangle$ and $\langle q, \sigma, s'_2\rangle$ in $\Delta$ for states $ s'_1$ and $s'_2$ such that $s_1 \sim_\A s'_1$ and $s_2 \sim_\A s'_2$. As $\A$ is nice, in particular, semantically deterministic, we have that $s'_1 \sim_\A s'_2$. Hence, as $s_1 \sim_\A s'_1$ and $s'_2 \sim_\A s_2$, we get by the transitivity of $\sim_\A$ that $s_1 \sim_\A s_2$. Then, Proposition~\ref{C and CE are equivalent} implies that $L(\A^{s_1}) = L(\A^{s_1}_{\E})$ and $L(\A^{s_2}) = L(\A^{s_2}_{\E})$, and so we get that $s_1 \sim_{\A_{\E}} s_2$. Thus, $\A_{\E}$ is semantically deterministic.
	\end{proof}

	Let $\C$ be a nice minimal GFG-tNCW equivalent to $\A$, and let $\hat{\E}$ be the
	set of all allowed transitions in $\C$ that are not transitions of $\C$.
	By Propositions~\ref{C and CE are equivalent} and~\ref{ae is nice}, we have that $\C_{\hat{\E}}$ is a nice minimal GFG-tNCW equivalent to $\A$. Below we argue that it is also $\alpha$-maximal.

	\begin{prop}
		Let $\C$ be a nice GFG-tNCW, and let $\hat{\E}$ be the set of all allowed transitions in $\C$ that are not transitions of $\C$. Then, $\C_{\hat{\E}}$  is $\alpha$-maximal.
	\end{prop}

	\begin{proof}
		Let $\C=\langle \Sigma, Q, q_0, \delta, \alpha \rangle$, and consider an allowed transition $\zug{q, \sigma, s} \in Q\times \Sigma \times Q$ in $\C_{\hat{\E}}$. We prove that $\zug{q, \sigma, s}$ is an allowed transition also in $\C$. Hence, it is in $\Delta_{\hat{\E}}$, and thus is a transition in $\C_{\hat{\E}}$.

		By the definition of allowed transitions, there is a state $s'\in Q$ with $s \sim_{\C_{\hat{\E}}} s'$ such that $s' \in \delta_{\hat{\E}}(q, \sigma)$. Proposition~\ref{C and CE are equivalent} implies that $L(\C^s) = L(\C^s_{\hat{\E}}) = L(\C^{s'}_{\hat{\E}}) =  L(\C^{s'})$, and thus $s \sim_\C s'$. Also, by the definition of $\delta_{\hat{\E}}$, there is a state $s''\in Q$ such that $s''\sim_ \C s'$ and $s'' \in \delta(q, \sigma)$. Therefore, as the transitivity of $\sim_\C$ implies that $s\sim_\C s''$, we have that $\zug{q, \sigma, s}$ is also an allowed transition in $\C$, and we are done.
	\end{proof}

	Since the relation $\sim$ can be calculated in polynomial time~\cite{HKR02,KS15}, and so checking if a triple in $Q \times \Sigma \times Q$ is an allowed transition can be done in polynomial time, then applying $\alpha$-maximization on top of the minimization construction described in Section~\ref{sc minimization}, is still polynomial. We can thus conclude with the following.

	\begin{thm}
		Every GFG-tNCW $\A$ can be canonized into a nice minimal $\alpha$-maximal GFG-tNCW in polynomial time.
	\end{thm}

\begin{figure}[htb]
		\begin{center}
			\includegraphics[width=.42\textwidth]{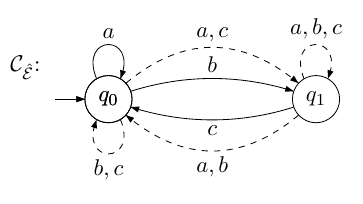}
			\captionof{figure}{The $\alpha$-maximal GFG-tNCW for the GFG-tNCW $\B_\S$ in Figure~\ref{B figure}.}%
			\label{alpha maximal fig}
		\end{center}
	\end{figure}

	\begin{exa}
		{\rm Consider the minimal GFG-tNCW $\B_\S$ appearing in Figure~\ref{B figure}. By applying $\alpha$-maximization to $\B_\S$, we obtain the $\alpha$-maximal GFG-tNCW $\C_{\hat{\E}}$ appearing in Figure~\ref{alpha maximal fig}.  \hfill \qed}
	\end{exa}

	\subsection{Canonicity in GFG-NCWs and tDCWs}%
	\label{sc canonicity determinisic}

	The positive canonicity results in Section~\ref{sec obtain} refer to GFG-tNCWs. In this section we study each of the features, namely GFGness and transition-based acceptance, in isolation. We start with GFG-NCWs. For automata with state-based acceptance, an analogue definition of isomorphism between automata $\A$ and $\B$ with acceptance conditions $\alpha_\A \subseteq Q_\A$ and $\alpha_\B \subseteq Q_\B$, seeks a bijection $\kappa: Q_\A \to Q_\B$ such that for every $q\in Q_\A$, we have that $q \in \alpha_\A$ iff $\kappa(q) \in \alpha_\B$, and for  every letter $\sigma \in \Sigma$, and state $q' \in Q_\A$, we have that $q' \in \delta_\A(q,\sigma)$ iff $\kappa(q') \in \delta_\B(\kappa(q),\sigma)$. It is easy to see that the DCWs $\A_1$ and $\A_2$ from Figure~\ref{2min dcws} are not isomorphic, which is a well known property of DCWs~\cite{Kup15}. In Theorem~\ref{no can gfg} below, we extend the ``no canonicity'' result to GFG-NCWs. As in the study of GFG-tNCWs, we focus on nice GFG-NCWs. Note that the definitions of the underlying properties of nice GFG-tNCWs naturally extend to GFG-NCWs.  For example, a GFG-NCW $\A$ is \emph{safe deterministic} if by removing $\alpha$-states and all transitions that touch them,  we remove all nondeterministic choices.  An exception is the definition of \emph{normal}. For a GFG-NCW $\A$, we say that $\A$ is normal if for all states $q \in Q$, if $q$ is not part of a cycle in $\A$, then $q \in \alpha$. Also, rather than $\alpha$-maximization, here we refer to \emph{$\alpha$-mutation}, which amounts to changing the membership of states in $\alpha$ while keeping the NCW nice and without changing its language.

	As we show in Theorem~\ref{no can gfg}, the fact the GFG automata we consider have transition-based acceptance conditions is crucial for canonization.

	\begin{thm}%
		\label{no can gfg}
		Nice, equivalent, and minimal GFG-NCWs need not be isomorphic, and need not be made isomorphic by $\alpha$-mutation.
	\end{thm}

	\begin{proof}
		Consider the language $L=(a+b)^* \cdot (a^\omega + b^\omega)$. In Figure~\ref{2min dcws}, we described
		two non-isomorphic DCWs $\A_1$ and $\A_2$ for $L$. The DCWs $\A_1$ and $\A_2$ can be viewed as nice GFG-NCWs.
		It is not hard to see that there is no $2$-state GFG-NCW for $L$, implying that $\A_1$ and $\A_2$ are nice, equivalent, and minimal GFG-NCWs that are not isomorphic, as required.
		 Moreover, no $\alpha$-mutation leads to isomorphism.
	\end{proof}

	We continue to tDCWs. Again, we refer to \emph{$\alpha$-mutation}, which amounts to changing the membership of transitions in $\alpha$ while keeping the tNCW nice and without changing its language.
	Note that unlike the case of GFG automata, where dualization, i.e., viewing a tNCW as a tNBW,
	need not result in an automaton for the complementary language, here results on tDCWs immediately apply also to tDBWs.
	We start with some bad news, showing that there is no canonicity also in the transition-based setting.
	Again, we refer to nice minimal automata. Note that some of the properties of nice GFG-tNCWs are trivial for minimal deterministic tDCW:\@ being minimal and deterministic, then clearly all states are reachable and GFG, the automata are semantically deterministic and safe deterministic, and we only have to make them normal by classifying transitions between safe components as $\alpha$-transitions. As we show in Theorem~\ref{tD not isomorphic}, the fact we consider GFG automata is crucial for canonization.

	\begin{thm}%
		\label{tD not isomorphic}
		Nice, equivalent, and minimal tDCWs and tDBWs need not be isomorphic, and need not be made isomorphic by $\alpha$-mutation.
	\end{thm}

\begin{figure}[htb]
			\begin{center}
				\includegraphics[width=.8\textwidth]{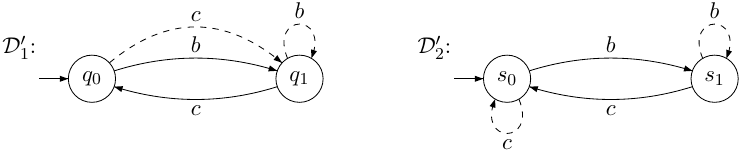}
				\captionof{figure}{Two non-isomorphic equivalent minimal nice tDCWs.}%
				\label{tdcw not iso min no a}
			\end{center}
		\end{figure}

	\begin{proof}
		As a first example, consider the GFG-tNCW $\B_\S$ from Figure~\ref{B figure}. Recall that $\B_\S$ is DBP;\@ that is, we can prune some of its transitions and get an equivalent tDCW\@. In Figure~\ref{tdcw not iso min} below, we describe two tDCWs obtained from it by two different prunnings. It is not hard to see that both tDCWs are nice and equivalent to $\B_\S$, yet are not isomorphic. 	 Moreover, no $\alpha$-mutation leads to isomorphism.

		\begin{figure}[htb]
			\begin{center}
				\includegraphics[width=.8\textwidth]{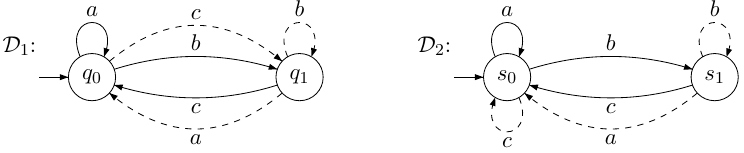}
				\captionof{figure}{Two non-isomorphic equivalent minimal nice tDCWs, obtained by different prunnings of~$\B_\S$.}%
				\label{tdcw not iso min}
			\end{center}
		\end{figure}

		By removing the $a$-labeled transitions from the tDCWs in Figure~\ref{tdcw not iso min}, we obtain a simpler example.  Consider the tDCWs $\D'_1$ and $\D'_2$ in Figure~\ref{tdcw not iso min no a}. It is easy to see that $L(\D'_1)=L(\D'_2)=(b+c)^* \cdot (b \cdot c)^\omega$. Clearly, there is no single-state tDCW for this language. Also, the tDCWs are not isomorphic, as a candidate bijection $\kappa$ has to be $\bar{\alpha}$-transition respecting, and thus have $\kappa(q_0)=s_0$ and $\kappa(q_1)=s_1$, yet then it is not $\alpha$-transition respecting. By dualizing the acceptance condition of $\D'_1$ and $\D'_2$, we obtain two non-isomorphic tDBWs for the complement language, of all words with infinitely many occurrences of $bb$ or $cc$. Also here,   no $\alpha$-mutation leads to isomorphism. 	\end{proof}

	Hence, both GFGness and the use of transition-based acceptance conditions are crucial for canonization, and we can conclude with the following.

	\begin{thm}
		There is no canonicity for minimal GFG-NCWs and for minimal tDCWs.
	\end{thm}

	\begin{rem}
	{\rm Note that our definition of $\alpha$-mutations does not add new transitions to the automaton. In particular, it preserves determinism and may thus seem weaker in the context of GFG-NCWs than the definition used for GFG-tNCWs. Note, however, that even if we allow the addition of transitions, there is no canonicity for GFG-NCWs. Indeed, we cannot add transitions to the DCWs $\A_1$ and $\A_2$  from Figure~\ref{2min dcws} and obtain a GFG-NCW that embodies both DCWs and still recognizes their language. As another example, as we discuss in Section~\ref{disc}, it is shown in~\cite{Sch20} that the NP-hardness proof of DBW minimization applies also to GFG-NBW\@. While GFG-NBWs and GFG-NCWs do not dualize each other, the proof in~\cite{Sch20} applies also for GFG-NCW minimization, and is based on associating a graph $G$ with a language $L_G$ such that  there is a correspondence between minimal GFG-NCWs for $L_G$ and minimal vertex covers for $G$. In more detail, for each vertex cover of $G$, the state space of a minimal GFG-NCW for $L_G$ has an $\bar{\alpha}$-state for each vertex of $G$ and an $\alpha$-state for each vertex in the cover. Thus, vertices in the cover induce two states each. The transitions of the automaton depend on the edges of $G$. Accordingly, different vertex covers induce GFG-NCWs with different structures. In particular, different minimal GFG-NCWs for $L_G$ are incomparable in many aspects, making $L_G$ resistant to more flexible definitions of canonicity. } \hfill \qed%
	\end{rem}

	The tDCWs in the proof of Theorem~\ref{tD not isomorphic} are safe isomorphic.
	We continue with some good news and study safe isomorphism between minimal nice tDCWs.

	We say that an $\omega$-regular language $L$ is \emph{GFG-helpful} if a minimal tDCW for $L$ is bigger than a minimal GFG-tNCW for $L$.
	\begin{thm}%
		\label{not safe iso}
		Consider an $\omega$-regular language $L$. If $L$ is not GFG-helpful, then
		every two nice and minimal tDCWs for $L$ are safe isomorphic.
	\end{thm}

	\begin{proof}
		Consider a language $L$ that is not GFG-helpful, and consider two nice minimal tDCWs $\A_1$ and $\A_2$ for $L$.
		Since $L$ is not GFG-helpful, then $\A_1$ and $\A_2$ are also nice minimal GFG-tNCWs for $L$. Hence, by Theorem~\ref{safe isomorphic}, they are safe isomorphic.
	\end{proof}

	Note that safe isomorphism for $\omega$-regular languages that are GFG-helpful is left open. Theorem~\ref{not safe iso} suggests that searching for a language $L$ that has two minimal tDCWs that are not safe isomorphic, we can restrict attention to languages that are GFG-helpful. Such languages are not common. Moreover, their canonicity is less crucial, as working with a minimal GFG-tNCW for them is more appealing.  An example of a family of languages that are GFG-helpful can be found in~\cite{KS15}, where it was shown that GFG-tNCWs may be exponentially more succinct than tDCWs. In Section~\ref{gh sec}, we demonstrate the execution of our algorithm on a GFG-helpful language in this family.

	\section{An Example}\label{gh sec}
	Recall that an $\omega$-regular language $L$ is GFG-helpful if a minimal tDCW for $L$ is bigger than a minimal GFG-tNCW for $L$. In this section, we present a GFG-helpful language, and demonstrate an execution of our minimization algorithm on a tDCW $\A$ that recognizes it.  The language belongs to the family of languages presented in~\cite{KS15}, where it was shown that GFG-tNCWs may be exponentially more succinct than tDCWs.

\begin{figure}[htb]
		\begin{center}
			\includegraphics[width=.5\textwidth]{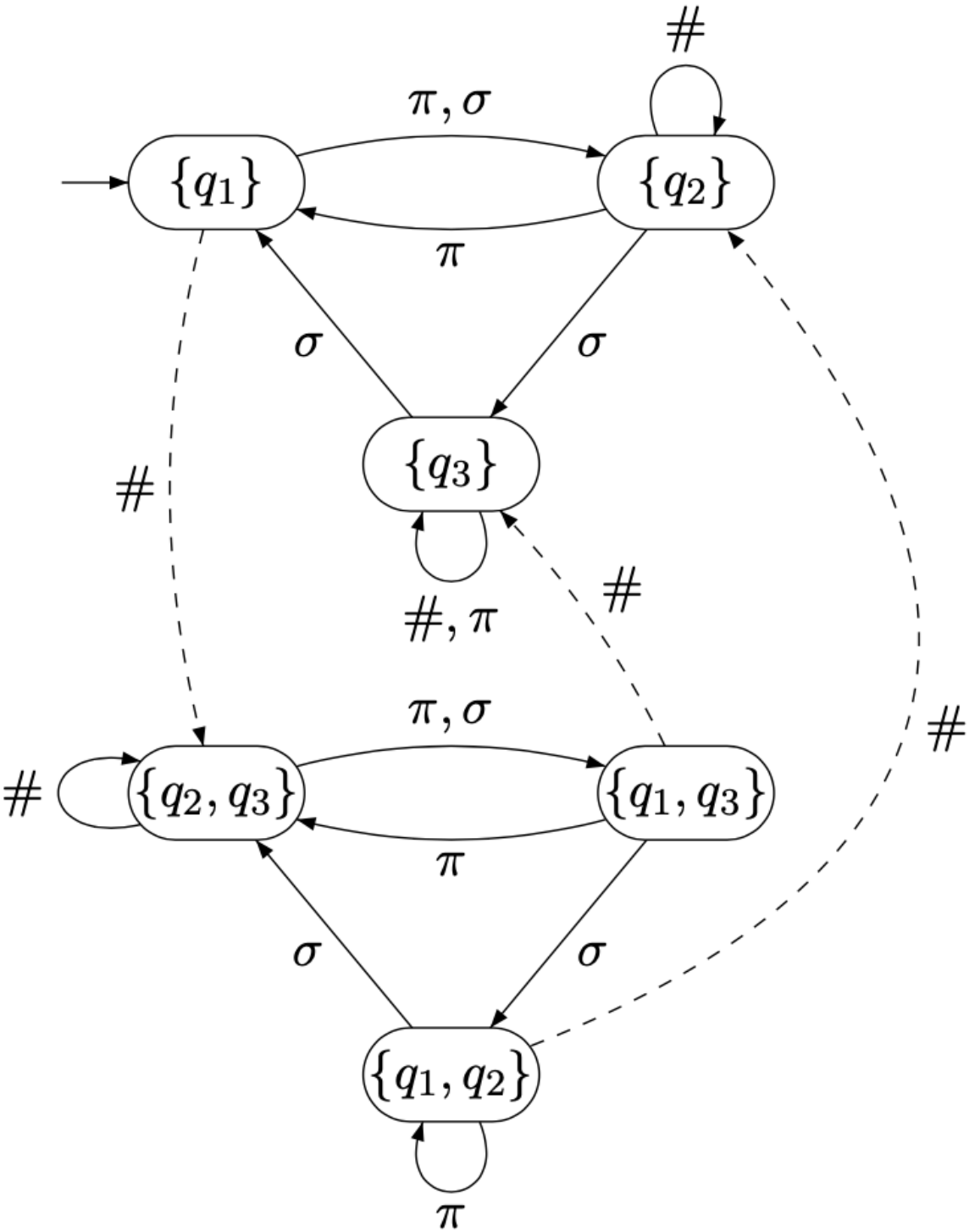}
			\captionof{figure}{The tDCW $\A$.}%
			\label{perm}
		\end{center}
	\end{figure}

	Consider the tDCW $\A$ over $\Sigma = \{ \sigma, \pi, \#\}$ that appears in Figure~\ref{perm}.

	Recall that dashed transitions are  $\alpha$-transitions. As $\A$ is normal and deterministic, it is nice.
	Let $L'=\bigcup_{i\in \{1, 2, 3\}} L_{{\it safe}}(\A^{\{q_i\}})$. Then, it is not hard to see that all the states in $\A$ are equivalent and they all recognizes the language $L(\A)= \{w : w \mbox{ has a suffix in }L'\}$.

	Intuitively, as the language of $\A$ can be specified in terms of the safe languages of the states $\{q_1\}, \{q_2\}$, and  $\{q_3\}$, it is not surprising that we can define a GFG-tNCW for the language on top of their safe component. The beauty of the example in~\cite{KS15} is the fact that a minimal tDCW for the language needs more states.

	As explained in~\cite{KS15}, the letters $\sigma$, $\pi$ and $\#$ can be viewed as actions performed to tokens placed on a graph with vertices $\{1, 2, 3\}$. The letter $\sigma$ moves each token from vertex $i$ to vertex $(i \mod 3)+1$. The letter $\pi$ switches the tokens in vertices $1$ and $2$ (and leaves the token in vertex $3$ in its vertex), and the letter $\#$ ``chops'' the path traversed by the token placed in vertex $1$ (and does not chop the paths traversed by the other tokens). Then, $\A$ recognizes the set of words in which at least one token traverses an infinite path.

	Clearly, a nondeterministic automaton for $L$ can guess a single token and follows its path, using the co-B\"uchi condition to make sure that there are only finitely many positions in which the token is in vertex $1$ and the letter $\#$ is read (and so the path is chopped). On the other hand, a deterministic automaton must remember the subset of tokens whose paths have not been chopped since some break-point. For example, when the tDCW $\A$ is in state $\{q_1,q_3\}$, indicating that the tokens in vertices $1$ and $3$ are ``alive'', and it reads the letter $\sigma$, it updates the locations of the two living tokens to $(1 \mod 3)+1$ and $(3 \mod 3)+1$, and so it moves to the state $\{q_1, q_2\}$. When, however, $\A$ reads the letter $\#$ from the state $\{q_1, q_3\}$, only the token in vertex $3$ stays alive, and $\A$ moves to the state $\{q_3\}$. A break-point happens when $\A$ reads $\#$ in the state $\{q_1\}$. Then, all paths have been chopped, and so $\A$ has an $\alpha$-transition to the state $\{q_2, q_3\}$, from which it continues to follow the tokens in vertices $2$ and $3$. The other $\alpha$-transitions in $\A$ can also be defined as safe transitions, but are in $\alpha$ due to normalization.

	As explained in~\cite{KS15} and demonstrated by our minimization algorithm, a GFG automaton for $L(\A)$ can also guess a single token and follow it. Essentially, a guess should be made when the path traversed by the token we follow is chopped. The GFG strategy makes use of the past and chooses to continue and follow the token with the longest path traversed so far.

\begin{figure}[htb]
		\begin{center}
			\includegraphics[width=.4\textwidth]{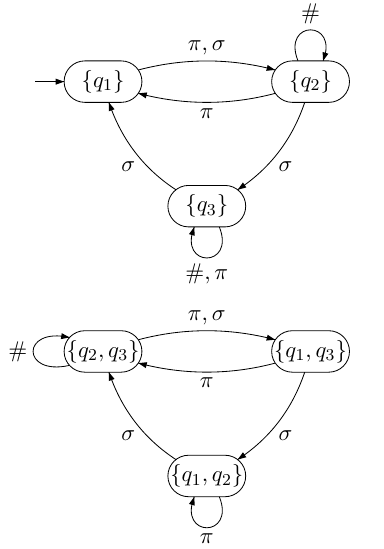}
			\captionof{figure}{The safe components of $\A$.}%
			\label{Exmpl safe components}
		\end{center}
	\end{figure}

	 Formally, we proceed with minimizing $\A$ by applying the minimization algorithm presented in Section~\ref{sc minimization}. %
	We begin by safe-centralizing $\A$, as explained in Section~\ref{sec sc}. First, by removing the $\alpha$-transitions of $\A$, we get the safe components described in Figure~\ref{Exmpl safe components}.

\begin{figure}[htb]
		\begin{center}
			\includegraphics[width=.35\textwidth]{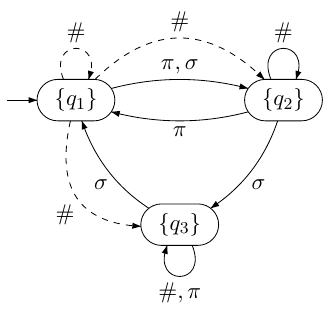}
			\captionof{figure}{The tNCW $\B_\S$, for $\S={\{\{ \{q_1\},\{q_2\}, \{q_3\}\}\}}$.}%
			\label{A2 figure}
		\end{center}
	\end{figure}

	Recall that $\A$ is nice and all of its states are equivalent. As $\{q_1, q_2\} \precsim \{q_1\}$, we have that $\A$ has a single frontier $\S=\{\{\{q_1\}, \{q_2\}, \{q_3\}\}\}$. The automaton $\B_{\S}$ appears in Figure~\ref{A2 figure}. As all the states of $\A$ are equivalent, we direct a $\#$-labeled $\alpha$-transition from $\{q_1\}$ to $\{q_1\},  \{q_2\}$ and to $\{q_3\}$.

	 We proceed by safe-minimizing $\B_\S$, as explained in Section~\ref{sec sd and sm}.
	 Note that the safe languages of the states $\{q_1\}, \{q_2\}$ and $\{q_3\}$ of the GFG-tNCW $\B_\S$ from Figure~\ref{A2 figure} are pairwise different. Indeed, $\#^\omega \in (L_{{\it safe}}(\A^{\{q_2\}})\cap L_{{\it safe}}(\A^{\{q_3\}}))\setminus L_{{\it safe}}(\A^{\{q_1\}})$ and $\sigma\cdot \#^\omega \in L_{{\it safe}}(\A^{\{q_2\}})\setminus L_{{\it safe}}(\A^{\{q_3\}})$. Thus, there are no strongly-equivalent states in $\B_\S$, and applying safe-minimization to $\B_\S$ results in the GFG-tNCW $\C$ identical to $\B_\S$.

To complete the example, we show that $L(\A)$ is GFG-helpful.

     \begin{prop}
     $L(\A)$ is GFG-helpful.
     \end{prop}

 	\begin{proof}
 	Assume by contradiction that there is a tDCW $\D=\zug{\Sigma,Q_\D,q^0_\D,\delta_\D,\alpha_\D}$ for $L(\A)$ with three states, and assume without loss of generality that $\D$ is normal (and thus is nice). As $\D$ is deterministic, it is GFG\@. Hence, by Theorem~\ref{safe isomorphic}, we have that $\D$ and the GFG t-NCW $\C$ are safe isomorphic. Therefore, $\D$ is obtained by pruning the transitions of the GFG-tNCW $\B_\S$ in Figure~\ref{A2 figure} and by choosing one of its states as the initial state.
 	Let $x_1$ be a finite word such that $\delta_\D(q^0_\D, x_1)=\{q_1\}$.
 	To reach a contradiction, we distinguish between three cases.
 	If the $\alpha$-transition $\langle \{q_1\}, \#, \{q_1\}\rangle$ is in $\D$, then consider the word $w = x_1\cdot \#^\omega$. As $\#^\omega \in L_{{\it safe}}(\A^{\{q_2\}})$, we conclude that $w \in L(\A)$. On the other hand, as $\#^\omega$ is rejected from the state $\{q_1\}$ in $\D$, we conclude that $w\notin L(\D)$.
 	Also, if the $\alpha$-transition $\langle \{q_1\}, \#, \{q_2\}\rangle$ is in $\D$, then consider the word $w = x_1 \cdot (\#\cdot \pi)^\omega$. As $(\#\cdot \pi)^\omega \in L_{{\it safe}}(\A^{\{q_3\}})$, we conclude that $w\in L(\A)$. On the other hand, as $(\#\cdot \pi)^\omega$ is rejected from the state $\{q_1\}$ in  $\D$, we conclude that $w\notin L(\D)$.
 	Finally, if the $\alpha$-transition $\langle \{q_1\}, \#, \{q_3\}\rangle$ is in $\D$, then consider the word $w = x_1 \cdot (\#\cdot \pi\cdot \sigma)^\omega$. As $(\#\cdot \pi\cdot \sigma)^\omega \in L_{{\it safe}}(\A^{\{q_2\}})$, we conclude that $w\in L(\A)$. On the other hand, as $(\#\cdot \pi\cdot \sigma)^\omega $ is rejected from the state $\{q_1\}$ in $\D$, we conclude that $w\notin L(\D)$.
 	As all different prunings of $\B_\S$ result in losing words from the language, we conclude that $\D$ cannot be equivalent to $\A$, and we have reached a  contradiction.
 	\end{proof}

	\section{Discussion}%
	\label{disc}

	Recall that for DCWs, minimization is NP-complete~\cite{Sch10}.
	Considering GFG-tNCWs rather than DCWs involves two modifications of the original DCW-minimization question: a transition to GFG rather than deterministic automata, and a transition to transition-based rather than state-based acceptance. A natural question that arises is whether both modifications are crucial for efficiency.
	It was recently shown~\cite{Sch20} that the NP-completeness proof of Schewe for DBW minimization can be generalized to GFG-NBWs and GFG-NCWs.
	This suggests that the consideration of transition-based acceptance has been crucial, and makes the study of tDBWs and tDCWs very appealing.

	The NP-hardness proof of DBW minimization fails for tDBWs, and it may well be that minimization of tDCWs (and hence, also tDBWs) can be solved in polynomial time. On the more sceptic side, our negative results about canonicity of tDCWs imply that no single quotient construction is possible.
	The use of transition-based acceptance is related to another open problem in the context of DBW minimization: is there a $2$-approximation polynomial algorithm for it; that is, one that generates a DBW that is at most twice as big as a minimal one.
	Note that a tight minimization for the transition-based case would imply a positive answer here. Indeed, given a DBW $\A$, one can view it as a  tDBW by considering all transitions that touch $\alpha$-states, as $\alpha$-transitions. Then, after minimizing the tDBW $\A$, one can efficiently turn it into an equivalent DBW that is at most twice as big. Note also that the vertex-cover problem, used in Schewe's reduction, has a polynomial $2$-approximation. As described in Section~\ref{intro}, there is recently growing use of automata with transition-based acceptance. Our work here is another evidence to their usefulness.

	We find the study of minimization of GFG automata of interest also beyond being an intermediate result in the quest for efficient transition-based DBW minimization. Indeed, GFG automata are important in practice, as they are used in synthesis and control, and in the case of the co-B\"uchi acceptance condition, they may be exponentially more succinct than their deterministic equivalences.
	Another open problem, which is interesting from both the theoretical and practical points of view, is minimization of GFG-tNBWs. Note that unlike the deterministic case, GFG-tNBWs and GFG-tNCWs are not dual. %
	Also, experience shows that GFGness has different theoretical properties for \buchi and co-\buchi automata~\cite{BKKS13, KS15,BKS17,BK18,AKL21}. For example, while GFG-NCWs are known to be exponentially more succinct than DCWs, the question of the succinctness of GFG-NBWs with respect to DBWs is still open, and even in case the answer is positive, succinctness is at most quadratic~\cite{KS15}. In addition, the arguments as to why a given GFG automaton is DBP or even \emph{almost DBP} (that is, can be determinized by pruning to an automaton that accepts exactly all words in the language with probability $1$) are different for B\"uchi and co-B\"uchi~\cite{AKL21}.

	\section{Glossary}%
	\label{app glos}

	All notations and definitions refer to a  GFG-tNCW $\A = \langle \Sigma, Q, q_0, \delta, \alpha  \rangle$.

	\subsection{Relations between states}
	\begin{itemize}[label=$\triangleright$]
		\item
		Two states $q,s\in Q$ are \emph{equivalent}, denoted $q \sim s$, if $L(\A^q) = L(\A^s)$.
		\item
		Two states $q,s\in Q$ are \emph{strongly-equivalent}, denoted $q \approx  s$, if $q \sim s$ and $L_{{\it safe}}(\A^q) = L_{{\it safe}}(\A^s)$.
		\item
		A state $q \in Q$ is \emph{subsafe-equivalent to} a state $s\in Q$, denoted $q\precsim s$, if $q \sim s$ and $L_{{\it safe}}(\A^q) \subseteq L_{{\it safe}}(\A^s)$.
	\end{itemize}

	\subsection{Properties of a GFG-tNCW}
	\begin{itemize}[label=$\triangleright$]
		\item
		$\A$ is \emph{semantically deterministic} if for every state $q\in Q$ and letter $\sigma \in \Sigma$, all the $\sigma$-successors of $q$ are equivalent: for every two states $s, s'\in \delta(q,\sigma)$, we have that $s \sim s'$.
		\item
		$\A$ is \emph{safe deterministic} if by removing its $\alpha$-transitions, we remove all nondeterministic choices. Thus,  for every state $q\in Q$ and letter $\sigma\in \Sigma$, it holds that $|\delta^{\bar{\alpha}}(q, \sigma)|\leq 1$.
		\item
		$\A$ is \emph{normal} if there are no $\bar{\alpha}$-transitions connecting different safe components. That is,
		for all states $q$ and $s$ of $\A$, if there is a path of $\bar{\alpha}$-transitions from $q$ to $s$, then there is also a path of $\bar{\alpha}$-transitions from $s$ to $q$.
		\item
		$\A$ is \emph{nice} if all the states in $\A$ are reachable and GFG, and $\A$ is normal, safe deterministic, and semantically deterministic.
		\item
		$\A$ is \emph{$\alpha$-homogenous} if for every state $q\in Q$ and letter $\sigma \in \Sigma$, either $\delta^\alpha(q, \sigma) =\emptyset$ or $\delta^{\bar{\alpha}}(q, \sigma) = \emptyset$.
		\item
		$\A$ is \emph{safe-minimal} if it has no strongly-equivalent states.
		\item
		$\A$ is \emph{safe-centralized} if for every two states $q, s\in Q$, if $q \precsim s$, then $q$ and $s$ are in the same safe component of $\A$.
	\end{itemize}

	\bibliographystyle{alphaurl}

	\bibliography{toSubmit}

	\appendix
\end{document}